\documentclass[aps,prd,onecolumn,eqsecnum,amsmath,nofootinbib,preprintnumbers,superscriptaddress]{revtex4-2}

\usepackage[dvipsnames]{xcolor}
\usepackage{color,graphicx,float,xcolor}
\usepackage{amsfonts,amssymb,theorem,mathrsfs,times}
\usepackage{bm}
\usepackage{multirow}
\usepackage{mathtools}
\usepackage{amsfonts,amssymb,theorem,mathrsfs}
\usepackage{dsfont}
\usepackage{setspace}
\usepackage{amsmath}
\makeatletter
\renewcommand{\aligned@a}[1][b]{\start@aligned{#1}\m@ne}
\makeatother
\let\oldgathered\gathered
\renewcommand{\gathered}[1][b]{\oldgathered[#1]}
\usepackage{graphicx}
\usepackage[colorlinks,linkcolor=blue,anchorcolor=blue,citecolor=blue]{hyperref}
\usepackage{ulem}
\usepackage{cancel}
\usepackage[justification = raggedright, singlelinecheck = true]{caption}
\usepackage{subcaption}
\usepackage{tikz}
\usetikzlibrary{decorations.pathmorphing,calc,arrows.meta,intersections}

\begin{document}
\title{
The entropy of black hole under second-order deviation from equilibrium}
\preprint{\hfill {\small {ICTS-USTC/PCFT-26-39}}}
\date{\today}

\author{Wen-Tao Fu}
\email{fuwentao2024@mail.ustc.edu.cn}

\affiliation{Interdisciplinary Center for Theoretical Study and Department of Modern Physics,\\
University of Science and Technology of China, Hefei, Anhui 230026, China}

\author{Ming-Fei Ji}
\email{jimingfei@mail.ustc.edu.cn}

\affiliation{Interdisciplinary Center for Theoretical Study and Department of Modern Physics,\\
University of Science and Technology of China, Hefei, Anhui 230026, China}

\author{Yu-Sen Zhou}
\email{zhou\_ys@mail.ustc.edu.cn}

\affiliation{Interdisciplinary Center for Theoretical Study and Department of Modern Physics,\\
University of Science and Technology of China, Hefei, Anhui 230026, China}

\author{Li-Ming Cao}
\email{caolm@ustc.edu.cn}
\affiliation{Interdisciplinary Center for Theoretical Study and Department of Modern Physics,\\
University of Science and Technology of China, Hefei, Anhui 230026, China}
\affiliation{Peng Huanwu Center for Fundamental Theory, Hefei, Anhui 230026, China}

\begin{abstract}
  We investigate the entropy of a dynamical black hole arising from second-order perturbations of a general stationary background with a bifurcate Killing horizon. Using Gaussian null coordinates, we study the geometry of the apparent horizon perturbatively up to second order. Within the covariant phase space formalism, to explore the contribution of matter fields, we introduce a new modified canonical energy, and establish a balance law relating the second-order variation of the entropy to the energy flux entering the black hole. We show that the entropy is given precisely by the area of the apparent horizon at second order when the null energy condition holds for the infalling matter, and that the variation of the entropy also obeys the second law. We also discuss the possibility that the area law continues to hold when the null energy condition is violated.
\end{abstract}

\maketitle

\section{Introduction}
\label{sec:introduction}

Black hole thermodynamics is one of the most profound developments in modern theoretical physics, as it unifies three originally independent frameworks: classical gravity, quantum theory, and statistical physics. Within this framework, a black hole behaves as a genuine thermodynamic system, equipped with temperature, entropy, and laws of evolution analogous to those of ordinary matter. This correspondence is made precise through the four laws of black hole mechanics, which establish a direct relation between horizon dynamics and thermodynamic behavior. Among these quantities, entropy plays a particularly central role: it not only governs the thermodynamic properties and stability of black holes, but also encodes microscopic information that is expected to shed light on the quantum structure of spacetime~\cite{Wald:1999vt}.

From a broader perspective, black hole entropy provides a unique window into several fundamental puzzles, including the statistical origin of gravitational degrees of freedom, the fate of information in gravitational collapse and evaporation, and the emergence of semiclassical geometry from an underlying quantum theory~\cite{Strominger:1996sh,Maldacena:1997re,Harlow:2014yka}. For this reason, understanding how entropy should be defined beyond strict stationarity is not merely a technical refinement, but a key step toward a deeper formulation of quantum gravity.

The earliest key insight into black hole entropy was provided by Bekenstein, who argued that the entropy of a black hole should be proportional to the area of its event horizon, $S_{\mathrm{BH}} \propto A$~\cite{Bekenstein:1973ur}. Subsequently, Bardeen, Carter, and Hawking formulated the four laws of black hole mechanics and established precise relations among mass, surface gravity, and horizon area~\cite{Bardeen:1973gs}. Hawking later demonstrated that black holes emit thermal radiation with temperature $T = \hbar \kappa/2\pi$~\cite{Hawking:1975vcx}, leading to the celebrated entropy formula $S = A/(4 G \hbar)$. These developments firmly established the thermodynamic interpretation of stationary black holes in general relativity.

Several decades later, Wald and collaborators developed the covariant phase space formalism in the context of gravity, in which black hole entropy is identified with a Noether charge associated with diffeomorphism invariance~\cite{Lee:1990nz,Wald:1990mme,Wald:1993nt,Iyer:1994ys}. This approach yields a general definition of black hole entropy for arbitrary diffeomorphism-invariant theories of gravity. However, this construction is intrinsically tied to stationary backgrounds, where a preferred Killing horizon exists. Extending the notion of entropy to genuinely dynamical black holes remains a longstanding challenge.

A significant recent development was made by Wald and collaborators~\cite{Hollands:2024vbe}. They use the covariant phase space formalism to construct a dynamical extension of black hole entropy, which is also the approach we adopted. Their analysis shows that, in general, the entropy is not associated with the area of the cross-section of the event horizon, but rather with the area of the apparent horizon, suggesting that the latter provides a more local and physically meaningful notion of entropy in nonstationary settings. Another approach is based on higher-curvature entropy functionals inspired by holographic entanglement entropy and the linearized second law, leading to the Dong-Wall entropy~\cite{Dong:2013qoa,Wall:2015raa}. Nevertheless, these results have been established mainly for first-order perturbations of stationary black holes~\cite{Rignon-Bret:2023fjq,Furugori:2025pmn}. Although there are some works on vacuum second-order perturbations~\cite{Hollands:2024vbe,Kong:2024sqc}, an explicit expression for the second-order correction to the entropy remains unavailable, and no satisfactory conclusion has yet been reached for second-order perturbations incorporating matter fields.

Recently, Ashtekar et al. developed a nonperturbative framework for black hole thermodynamics far from equilibrium, suggesting that the entropy is also proportional to the area of the marginally trapped surfaces associated with quasi-local horizons~\cite{Ashtekar:2025qqa,Ashtekar:2026jdz}. Their approach is based on quasi-local notions of horizons~\cite{Ashtekar_2004} and relies on the way in which the thermodynamic quantities of black holes are identified. This is different from the approaches based on the covariant phase space analysis.

In this work, within the covariant phase space formalism, we further investigate the structure of dynamical black hole entropy by considering second-order perturbations of a general stationary spacetime. Working in Gaussian null coordinates, we perturbatively construct the apparent horizon and analyze the associated expansions and area element. By combining these geometric results with the covariant phase space formalism, we derive the modified canonical energy and relate it to the second-order variation of entropy. Our results show that, at second order, the entropy remains proportional to the area of the apparent horizon under the null energy condition, whereas additional conditions are required to establish the same proportionality without imposing the null energy condition. This result partially extends the proposal of~\cite{Hollands:2024vbe} beyond linear order and supports the interpretation that the apparent horizon plays a distinguished role in characterizing dynamical black hole entropy.

The rest of this paper is organized as follows. In Section~\ref{sec:geometry_and_horizon}, we formulate the perturbed near-horizon geometry in Gaussian null coordinates, fix the gauge conditions, and construct the apparent horizon as a deformation of the event horizon. We also compute the corresponding null normals, expansions, and area element to the order needed for the entropy comparison. In Section~\ref{sec:modified_canonical_energy}, we review the covariant phase space formalism and express the second-order entropy variation in terms of the modified canonical energy and the constraint contribution on the horizon. In Section~\ref{sec:dynamical_entropy}, we combine these ingredients with the Raychaudhuri equation, the shear term, the boundary condition, and the consequence of the null energy condition to show how the quadratic contributions cancel, thereby relating the entropy to the apparent-horizon area. We conclude with a summary of the main result and its implications in Section~\ref{sec:conclusion}.

\section{Black Hole Geometry and Apparent Horizon}
\label{sec:geometry_and_horizon}
In this section, we set up the near-horizon geometry in Gaussian null coordinates and fix the gauge conditions used in the perturbative analysis. We then construct the apparent horizon as a deformation of the event horizon, calculate the relevant null expansions, and derive the second-order expansion of its area element. These geometric results provide the ingredients needed to compare the apparent-horizon area with the entropy in later sections.

\subsection[Geometric Setup]{Geometric Setup}
We consider a dynamical black hole perturbed from a stationary black hole background $(M,g)$ in $n$-dimensional spacetime. The event horizon of the stationary black hole is taken to be a bifurcate Killing horizon $\mathcal{H}$, where we denote the future horizon as $\mathcal{H}^+$, the past horizon as $\mathcal{H}^-$, and the codimension-$2$ bifurcation surface as $\mathcal{B}$. The background spacetime is assumed to be stationary, possessing a Killing vector field $\xi^a$ which generates the Killing horizons~\cite{Visser:2024pwz}.

In the stationary case, near $\mathcal{H}^+$, the spacetime metric in affinely parametrized Gaussian null coordinates $\{u,v,x\}$ with $x=(x^1,x^2,\ldots,x^{n-2})$ is given by
\begin{equation}
\begin{aligned}
    g_{ab}(u,v,x)
  =& -(du)_a (dv)_b
  -(du)_b (dv)_a
  - u^2 \alpha(u,v,x) (dv)_a (dv)_b
  - u \beta_i(u,v,x) (dv)_a (dx^i)_b \\
  &- u \beta_i(u,v,x) (dv)_b (dx^i)_a
  + \gamma_{ij}(u,v,x)\,(dx^i)_a (dx^j)_b \, ,
\end{aligned}
\end{equation}
where the $u=0$ hypersurface corresponds to $\mathcal{H}^+$, and $v$ is chosen as an affine parameter along the null generators, the functions $\alpha(u,v)$, $\beta_i(u,v)$ and $\gamma_{ij}(u,v,x)$ are assumed to be finite at $u=0$~\cite{Jia:2025tgf}. The inverse metric is
\begin{eqnarray}
  g^{ab}(u,v,x)
  &=& u^2[\alpha(u,v,x) + \beta^2](\partial_u)^a(\partial_u)^b
  -(\partial_u)^a(\partial_v)^b-(\partial_u)^b(\partial_v)^a-u\beta^i(u,v,x)(\partial_i)^a(\partial_u)^b\nonumber\\
  &&
  -u\beta^i(u,v,x)(\partial_i)^b(\partial_u)^a
  +\gamma^{ij}(u,v,x)(\partial_i)^a(\partial_j)^b \, ,
\end{eqnarray}
where $\gamma^{ij}$ denotes the inverse of $\gamma_{ij}$, $\beta^{i}=\gamma^{ij}\beta_j$, and $\beta^2=\beta_i\beta^i$. The metric can be decomposed as
\begin{equation}
  g_{ab} \overset{\mathcal{H}^+}{=} -2 k_{(a} l_{b)} + \gamma_{ab} \, ,
\end{equation}
here, $\overset{\mathcal{H}^+}{=}$ denotes equality evaluated on $\mathcal{H}^+$, and the transverse metric can be expressed as
\begin{equation}
  \gamma_{ab}
  = \gamma_{ij}\,(dx^i)_a (dx^j)_b
  \overset{\mathcal{H}^+}{=} \gamma_{ij}\,m^i_a m^j_b \, ,
\end{equation}
where
\begin{equation}
 k^a := (\partial_v)^a\, ,\quad
l^a := (\partial_u)^a\, ,\quad
m_i^a := (\partial_i)^a\, ,
\end{equation}
and
\begin{equation}
m^i_a:= \gamma^{ij}(u,v,x)m_j^b g_{ab}
  = (dx^i)_a - u\beta^i(u,v,x)(dv)_a
  \overset{\mathcal{H}^+}{=} (dx^i)_a \, .
\end{equation}
It is easy to check that they satisfy that
\begin{equation}
k_al^a = -1\, ,\quad k_am_i^a\overset{\mathcal{H}^+}{=}0\, ,\quad l_am_i^a=0\, .
\end{equation}
Note that
\begin{equation}
  k_a
  = g_{ab}k^b
  = -(du)_a - u^2\alpha(u,v,x)\,(dv)_a - u\beta_i(u,v,x)\,(dx^i)_a
  \overset{\mathcal{H}^+}{=} -(du)_a \, , \qquad
  l_a
  = g_{ab}l^b
  = -(dv)_a \, .
\end{equation} 

Below we demonstrate how to decompose $\xi^a$ in terms of the null basis $(k^a,l^a)$. Since $\xi^a$ is normal to $\mathcal{H}$, it is tangent to the null geodesic generators of $\mathcal{H}$. Along $\mathcal{H}^+$ it is therefore proportional to $k^a$, whereas along $\mathcal{H}^-$, it is proportional to $-l^a$, because $\xi^a$ is past directed on $\mathcal{H}^-$ whereas $l^a$ is future directed. On $\mathcal{H}^+$, the Killing vector $\xi^a$ obeys $\xi^b\nabla_b\xi^a \overset{\mathcal{H}^+}{=} \kappa\xi^a$ and the outgoing null normal is affinely parameterized, $k^b\nabla_b k^a \overset{\mathcal{H}^+}{=} 0$. Correspondingly, on $\mathcal{H}^-$, one similarly has $\xi^a\nabla_a\xi^b \overset{\mathcal{H}^-}{=} -\kappa\xi^b$ and $l^a\nabla_a l^b=0$. Therefore, one can choose a combination of $k^a$ and $l^a$ such that~\cite{Visser:2024pwz}
\begin{equation}
  \xi^a = \kappa \left( v k^a - u l^a \right) \, ,
\end{equation}
where $\kappa$ is the surface gravity. This ensures that $\xi^a$ acts like a local Lorentz boost near the horizon and respects the fact that $\xi^a=0$ at $\mathcal{B}$, where $u = v = 0$. Therefore, the isometry generated by $\xi^a$ restricts the form of the metric functions $\alpha$, $\beta_i$ and $\gamma_{ij}$ so that the dependence on $u$ and $v$ occurs only through the combination $\kappa u v$~\cite{Hollands:2022fkn,Bhattacharyya:2021jhr}:
\begin{equation}
  \alpha = \alpha (\kappa u v, x) , \qquad
  \beta_i = \beta_i(\kappa u v, x) , \qquad
  \gamma_{ij} = \gamma_{ij}(\kappa u v, x) \, .
\end{equation}

We now proceed with the perturbative analysis and first give the definition of the metric perturbation. We introduce a one-parameter family of metrics $g_{ab}(s)$ labeled by $s$.
\begin{equation}
  g_{ab}(s) = g_{ab} + s \delta g_{ab} + \frac{1}{2} s^2 \delta^2 g_{ab} + \mathcal{O}(s^3) \, .
\end{equation}
The validity of the approximation, upon setting $s = 1$, depends on the smallness of the perturbation $\delta g_{ab}$ and $\delta^2 g_{ab}$, i.e.,
\begin{equation}
  \delta g_{ab} := \left.\frac{d}{ds} g_{ab}(s)\right|_{s=0} \, ,
  \qquad
  \delta^2 g_{ab} := \left.\frac{d^2}{ds^2} g_{ab}(s)\right|_{s=0} \, .
\end{equation}
To proceed with the perturbative analysis, we use the diffeomorphism gauge freedom to fix the spacetime point identification between the unperturbed and perturbed configurations. This leads us to introduce the following gauge conditions to simplify the perturbative analysis~\cite{Visser:2024pwz}.
\begin{enumerate}
  \item[(1)]
    The event horizon of the perturbed black hole is identified with the Killing horizon of the unperturbed background. Moreover, $\mathcal{H}^+$ and $\mathcal{H}^-$ remain at $u = 0$ and $v = 0$, respectively, after the perturbation.

  \item[(2)]
    The null vectors $k^a$ and $l^a$ are held fixed under the perturbation,
    \begin{equation}
      \delta^m k^a = 0 \, , \qquad
      \delta^m l^a = 0 \, , \qquad
      (m=1,2)\, .
    \end{equation}
    Additionally, $k^a$ remains null only on $\mathcal{H}^+$ and normal to $\mathcal{H}^+$, while $l^a$ stays null everywhere under the perturbation. Combined with the requirement $\delta^m(k^a l_a)=0$, this implies the following conditions
    \begin{equation}
      k^a \delta^m g_{ab} \overset{\mathcal{H}^+}{=} 0 \, , \qquad
      l^a \delta^m g_{ab} = 0 \, , \qquad
      (m=1,2)\, .
      \label{gauge_condition_kl}
    \end{equation}
    We further require that $k^a$ remains affinely parametrized on $\mathcal{H}^+$ and $l^a$ remains affinely parametrized throughout the entire spacetime under the perturbation,
    \begin{equation}
      \delta^m \!\left( k^b
      \nabla_b k^a \right) \overset{\mathcal{H}^+}{=} 0 \, , \qquad
      \delta^m \!\left( l^b
      \nabla_b l^a \right) = 0 \, , \qquad
      (m=1,2)\, .
    \end{equation}

  \item[(3)]
    The Killing field $\xi^a$ continues to be null and tangent to the geodesic generators of the perturbed horizon.
    \begin{equation}
    \delta^m(g_{ab}\xi^a\xi^b)\overset{\mathcal{H}}{=}0 \, , \qquad \zeta_a\delta^m\xi^a\overset{\mathcal{H}}{=}0 \, , \qquad (m=1,2)\, ,
    \end{equation}
    where $\zeta^a$ is an arbitrary spacelike vector orthogonal to both $k^a$ and $l^a$. Together with condition~\eqref{gauge_condition_kl}, $\xi^a\delta^m g_{ab}\overset{\mathcal{H}}{=}0$, this means on the future horizon $\delta^m\xi^a$ is proportional to $k^a$ and on the past horizon to $l^a$.
\end{enumerate}

From the above gauge conditions, it follows that the affine parameters $v$ and $u$ are fixed under the perturbation. We will also fix the spatial coordinates $x^i$. Therefore, the perturbation of a stationary background affects only the metric functions, which we expand as
\begin{align}
  \alpha(s;u,v,x)
  &= \alpha(\kappa u v,x) + s\delta\alpha(u,v,x) + \frac{1}{2} s^2\delta^2\alpha(u,v,x) \, , \\
  \beta_i(s;u, v, x)
  &= \beta_i(\kappa u v,x) + s\delta \beta_i(u,v,x) + \frac{1}{2} s^2\delta^2\beta_i(u,v,x) \, , \\
  \gamma_{ij}(s;u,v,x)
  &= \gamma_{ij}(\kappa u v,x) + s \delta\gamma_{ij}(u,v,x) + \frac{1}{2} s^2 \delta^2\gamma_{ij}(u,v,x) \, .
\end{align}
Thus, after introducing the perturbation, the metric takes the following form
\begin{equation}
\begin{aligned}
g_{ab}(s;u,v,x)
  =& -(du)_a (dv)_b
  -(du)_b (dv)_a
  - u^2 \alpha(s;u,v,x) (dv)_a (dv)_b
  - u \beta_i(s;u,v,x) (dv)_a (dx^i)_b \\
  &- u \beta_i(s;u,v,x) (dv)_b (dx^i)_a
  + \gamma_{ij}(s;u,v,x)\,(dx^i)_a (dx^j)_b \, , 
\end{aligned}
\label{eq:metric}
\end{equation}
which is the foundation of all the following discussions.
\begin{figure}[htbp]
    \centering
\begin{tikzpicture}[
  scale=0.1, 
  line cap=round,
  line join=round,
  >=Latex
]

\draw[very thick,-Stealth,name path=kline] (-5,-5) -- (30,30) node[right] {$k^{a}$};
\draw[very thick,-Stealth,name path=lline] (5,-5) -- (-30,30) node[left] {$l^{a}$};

\draw[thick,dashed,name path=dashedline] (18,7) -- (-2,27);

\draw[very thick] (-7,13) -- (-5,15);

\def\k{1}

\pgfmathsetmacro{\xL}{-5}
\pgfmathsetmacro{\yL}{15}
\pgfmathsetmacro{\xR}{25}
\pgfmathsetmacro{\yR}{25}

\pgfmathsetmacro{\U}{rad(atan(\k))}
\pgfmathsetmacro{\V}{\k/(1+\k*\k)}

\pgfmathsetmacro{\xm}{(\xL+\xR)/2}
\pgfmathsetmacro{\sx}{(\xR-\xL)/2}

\pgfmathsetmacro{\ym}{(\yL+\yR)/2}

\pgfmathsetmacro{\b}{((\yR-\yL)-2*\sx)/(2*(\U-\V))}
\pgfmathsetmacro{\a}{\sx-\b*\V}

\draw[very thick,name path=curve,domain=-1:1,samples=200,smooth,variable=\t]
  plot ({\xm+\sx*\t},{\ym+\a*\t+\b*rad(atan(\k*\t))});

\path[name intersections={of=dashedline and kline, by=Pk}];
\path[name intersections={of=dashedline and curve, by=Pc}];

\fill (Pk) circle (0.6);
\fill (Pc) circle (0.6);
\fill (0,0) circle (0.6);

\node at (-25,20) {\small $\mathcal{H}^{-}$};
\node at (27,21) {\small $\mathcal{H}^{+}$};
\node at (5,0) {\small $\mathcal{B}$};
\node at (7,24) {\small $\mathcal{T}(v)$};
\node at (18,13) {\small $\mathcal{C}(v)$};
\node at (26,31) {\small $v$};
\node at (-26,31) {\small $u$};
\node at (-7,20) {\small $U(v,x)$};

\end{tikzpicture}
    \caption{A schematic figure for the apparent horizon in the perturbed black hole spacetime. Before the matter field enters, the apparent horizon is null and distinct from the event horizon. Since $k^a$ need not be null outside the event horizon, curves drawn parallel to it need not be null; this segment is therefore only schematic.}
    \label{fig:schematic}
\end{figure}
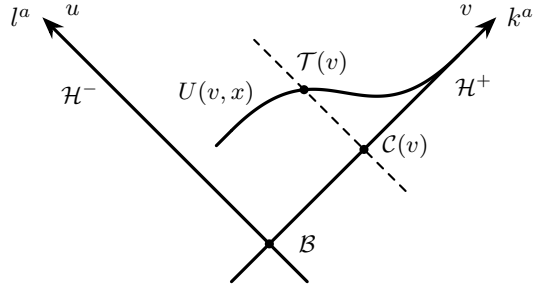

\subsection{Geometric Quantities on the Apparent Horizon}

The apparent horizon is defined as a marginally outer trapped surface on which the expansion of the outgoing null congruence vanishes, while that of the ingoing null congruence remains negative. Since the expansion is defined locally, the apparent horizon provides a more local notion of the black-hole boundary than the event horizon, thereby avoiding the teleological nature of the event horizon in discussions of local physics~\cite{Ashtekar_2004}. Therefore, below we study its relation to dynamical black-hole entropy by analyzing the area of the apparent horizon.

In the stationary case, the apparent horizon coincides with the event horizon. The outgoing null expansion is then given by $\theta_v=\nabla_a k^a$ on $\mathcal H^+$, which describes the rate of change of the area element $dA$ along the affine parameter $v$:
\begin{equation}
  \theta_v\,dA=\frac{d}{dv}(dA)\,.
  \label{eq:theta_v_area_element}
\end{equation}
Here $dA=\sqrt{\det\gamma}\,d^{n-2}x$ is the area element on a $(n-2)$-dimensional horizon cross section. From Eq.~\eqref{eq:theta_v_area_element}, we obtain
\begin{equation}
  \theta_v=\frac12\gamma^{ij}\partial_v\gamma_{ij}
  =\partial_v\ln\sqrt{\det\gamma}\,.
\end{equation}
The position of the apparent horizon $\mathcal A$ is denoted by~\cite{Visser:2024pwz}
\begin{equation}
  u=U(s;v,x)\,.
  \label{apparent horizon location}
\end{equation}
Here $U(s;v,x)$ describes the affine null distance away from the event horizon and admits the following expansion under perturbations:
\begin{equation}
  U(s;v,x)=s\delta U(v,x)
  +\frac{s^2}{2}\delta^2U(v,x)+\mathcal O(s^3)\,.
  \label{expansion of U}
\end{equation}
The position of the apparent horizon is illustrated in Fig.~\ref{fig:schematic}. Using Eq.~\eqref{expansion of U}, an arbitrary function $F(s;U(s;v,x),v,x)$ on the apparent horizon admits a second-order perturbative expansion about its unperturbed value on the event horizon:
\begin{equation}
\begin{aligned}
  F(s;U(s;v,x),v,x)
  &=F(0,v,x)+s\delta F(0,v,x)
  +\frac{s^2}{2}\delta^2F(0,v,x)\\
  &\quad+s\delta U(v,x)\partial_uF(0,v,x)
  +\frac{s^2}{2}[\delta U(v,x)]^2\partial_u^2F(0,v,x)\\
  &\quad+s^2\delta U(v,x)\partial_u\delta F(0,v,x)
  +\frac{s^2}{2}\delta^2U(v,x)\partial_uF(0,v,x)
  +\mathcal O(s^3)\,.
\end{aligned}
\label{eq:expansion_scheme}
\end{equation}
Here $F(0,v,x)$ denotes $F(0;0,v,x)$, $\delta F(0,v,x)=\partial_sF(0;0,v,x)$, and similar notation is used for higher orders.

To study the geometry of the apparent horizon, we consider its cross section $\mathcal T(v)$ at fixed $v$, which is a future marginally outer trapped surface. The tangent vectors and induced metric are
\begin{equation}
  e_i^a:=m_i^a+\partial_iU\,l^a,\qquad
  g_{ab}(s)e_i^ae_j^b=\gamma_{ij}(s;U,v,x)\,.
\label{eq:apparent_horizon_tangent_vectors}
\end{equation}
The ingoing null normal covector on $\mathcal T(v)$ is given by
\begin{equation}
  \tilde l_a\overset{\mathcal T}{=}l_a=-(dv)_a\,,
\end{equation}
whose dual vector is
\begin{equation}
  \tilde l^a\overset{\mathcal T}{=}
  g^{ab}\tilde l_b=(\partial_u)^a=l^a\,.
\end{equation}
The outgoing null normal covector $\tilde k_a$ is constructed from two linearly independent normal covectors on the codimension-two surface $\mathcal T(v)$, which are given by
\begin{equation}
  (df_1)_a=(du)_a-\partial_vU(dv)_a-\partial_iU(dx^i)_a\,,
\end{equation}
and
\begin{equation}
  (df_2)_a=(dv)_a\,,
\end{equation}
where $f_1=u-U(s;v,x)$ and $f_2=v$. We construct $\tilde k_a$ as a linear combination of $(df_1)_a$ and $(df_2)_a$, with the coefficients fixed by $\tilde k^a\tilde l_a=-1$ and $\tilde k^a\tilde k_a=\mathcal O(s^3)$. According to Eq.~\eqref{eq:expansion_scheme}, this gives, on $\mathcal T(v)$,
\begin{equation}
\begin{aligned}
  \tilde k_a\overset{\mathcal T}{=}
  -(du)_a-\frac{s^2}{2}
  \left[\lambda(\delta U)^2+(D\delta U)^2
  +2\beta^i\delta U D_i\delta U\right](dv)_a+\left(sD_i\delta U+\frac{s^2}{2}D_i\delta^2U\right)(dx^i)_a
  +\mathcal O(s^3)\,,
\end{aligned}
\end{equation}
with $\lambda=\alpha+\beta^2$. The corresponding dual vector is
\begin{equation}
\begin{aligned}
  \tilde k^a\overset{\mathcal T}{=}
  &\,k^a+\frac{s^2}{2}\left[(D\delta U)^2-\lambda(\delta U)^2\right]l^a+\Bigl [s(D^i\delta U+\beta^i\delta U)+\frac{s^2}{2}(D^i\delta^2U+\beta^i\delta^2U)\\
  &+s^2(\delta\gamma^{ij}D_j\delta U+\delta\beta^i\delta U)+s^2\delta U(\partial_u\gamma^{ij}D_j\delta U+\delta U\partial_u\beta^i)\Bigr ]m_i^a+\mathcal O(s^3)\,.
\end{aligned}
\end{equation}
Here, $D_i$ represents the codimension-2 intrinsic covariant derivative, $D^i=\gamma^{ij}D_j$, and $(D\delta U)^2=(D_i\delta U)(D^i\delta U)$. The metric and projection operator on $\mathcal T(v)$ are
\begin{equation}
\begin{aligned}
g_{ab}(s)&\overset{\mathcal T}{=}
-2\tilde k_{(a}\tilde l_{b)}
+\gamma_{ij}(s;U,v,x)e^i_ae^j_b,\\
\gamma^a{}_b&\overset{\mathcal T}{=}
\delta^a{}_b+\tilde k^a\tilde l_b+\tilde l^a\tilde k_b
=e_i^ae_b^i\,,
\end{aligned}
\end{equation}
where
\begin{equation}
e^i_a:=\gamma^{ij}(s;U,v,x)g_{ab}(s)e_j^b\,.
\end{equation}
Using the first relation in Eq.~\eqref{eq:apparent_horizon_tangent_vectors} and projection operator, we express the expansion $\tilde{\theta}_{\tilde{k}}$ on $\mathcal{T}(v)$ as
\begin{equation}
\begin{aligned}
  \tilde\theta_{\tilde k}
  &=\gamma^a{}_b\nabla_a\tilde k^b
  =e_i^ae_b^i\nabla_a\tilde k^b\\
  &=\theta_v
  +s\left(\delta\theta_v+\delta U\,\partial_u\theta_v\right)+\frac{s^2}{2}\Bigl\{
  \delta^2\theta_v+\delta^2U\,\partial_u\theta_v
  +2\delta U\,\partial_u\delta\theta_v
  +(\delta U)^2\partial_u^2\theta_v
  +\left[(D\delta U)^2-\lambda(\delta U)^2\right]\theta_u
  \Bigr\}\\
  &\quad+D_i\Biggl[
  D^i\left(s\delta U+\frac{s^2}{2}\delta^2U\right)
  +s\beta^i\delta U+\frac{s^2}{2}\beta^i\delta^2U
  +s^2\left(\delta\gamma^{ij}D_j\delta U
  +\delta\beta^i\delta U\right)
  +s^2\delta U\left(
  \partial_u\gamma^{ij}D_j\delta U
  +\delta U\,\partial_u\beta^i\right)
  \Biggr]\\
  &\quad+\frac{s^2}{2}
  (D^i\delta U+\beta^i\delta U)
  D_i(\gamma^{jk}\delta\gamma_{jk})
  +s^2\delta U(D^i\delta U+\beta^i\delta U)
  D_i\theta_u
  +\mathcal O(s^3)\,.
\end{aligned}
\label{eq:apparent_horizon_expansion_exact}
\end{equation}
Next, we impose the defining condition for an apparent horizon, namely that the expansion $\tilde{\theta}_{\tilde{k}}$ must vanish which implies
\begin{equation}
\begin{aligned}
  &\theta_v=0,\\
  &\delta\theta_v+\delta U\partial_u\theta_v
  +D_i(D^i\delta U+\beta^i\delta U)=0,\\
  &\delta^2\theta_v+\delta^2U\partial_u\theta_v
  +(\delta U)^2(\partial_u^2\theta_v-\lambda\theta_u)
  +2\delta U\partial_u\delta\theta_v
  +(D\delta U)^2\theta_u\\
  &+D_i\Bigl[
  D^i\delta^2U+\beta^i\delta^2U
  +2(\delta\gamma^{ij}D_j\delta U+\delta\beta^i\delta U)
  +2\delta U(\partial_u\gamma^{ij}D_j\delta U
  +\delta U\partial_u\beta^i)\Bigr]\\
  &
  +(D^i\delta U+\beta^i\delta U)
  D_i(\gamma^{jk}\delta\gamma_{jk})
  +2\delta U(D^i\delta U+\beta^i\delta U)D_i\theta_u=0\,.
\end{aligned}
\label{eq:apparent_horizon_condition}
\end{equation}

Below, we introduce two relations to simplify the second and third equalities in Eq.~\eqref{eq:apparent_horizon_condition}. At zeroth order, $\partial_u\theta_v=\partial_v\theta_u$~\cite{Visser:2024pwz}, while
\begin{equation}
  \theta_u(0,v,x)=-v\mu(x),
  \label{eq:theta_u}
\end{equation}
where $\mu(x)>0$~\cite{Jia:2025tgf}. This sign follows from the fact that the cross-section $\mathcal C(v)$ is a marginally outer trapped surface for $v>0$: the outgoing null expansion vanishes, whereas the ingoing null expansion is negative. Substituting Eq.~\eqref{eq:theta_u} into Eq.~\eqref{eq:apparent_horizon_condition}, we obtain
\begin{equation}
\begin{aligned}
  \delta\theta_v
  &=\mu\delta U-D_i(D^i\delta U+\beta^i\delta U),\\
  \delta^2\theta_v
  &=\mu\delta^2U
  -(\delta U)^2\partial_u^2\theta_v-2\delta U\partial_u\delta\theta_v-(D\delta U)^2\theta_u-\frac{\lambda v}{\mu}\left[\delta\theta_v+D_i(D^i\delta U+\beta^i\delta U)\right]^2\\
  &\quad-D_i\Bigl[D^i\delta^2U+\beta^i\delta^2U+2(\delta\gamma^{ij}D_j\delta U+\delta\beta^i\delta U)+2\delta U(\partial_u\gamma^{ij}D_j\delta U+\delta U\partial_u\beta^i)\Bigr]\\
  &\quad-(D^i\delta U+\beta^i\delta U)D_i(\gamma^{jk}\delta\gamma_{jk})-2\delta U(D^i\delta U+\beta^i\delta U)D_i\theta_u\,.
\end{aligned}
\label{eq:theta_relations}
\end{equation}

Finally, we expand the apparent-horizon area element $dA_{\mathcal T}$ at affine parameter $v$ around the event-horizon cross section at the same $v$, following the same procedure as in Eq.~\eqref{eq:expansion_scheme}:
\begin{equation}
\begin{aligned}
  dA_{\mathcal T}(s;U(s;v,x),v,x)
  &=dA_\epsilon(0,v,x)
  +s\delta dA_\epsilon(0,v,x)
  +\frac{s^2}{2}\delta^2dA_\epsilon(0,v,x)\\
  &\quad+s\delta U(v,x)\partial_udA_\epsilon(0,v,x)
  +\frac{s^2}{2}[\delta U(v,x)]^2
  \partial_u^2dA_\epsilon(0,v,x)\\
  &\quad+s^2\delta U(v,x)\partial_u\delta dA_\epsilon(0,v,x)
  +\frac{s^2}{2}\delta^2U(v,x)
  \partial_udA_\epsilon(0,v,x)
  +\mathcal O(s^3)\,,
\end{aligned}
\label{eq:apparent_horizon_area_expansion_second_order1}
\end{equation}
where $dA_\epsilon(0,v,x)$ denotes the area element of the cross section $\mathcal C(v)$ of the event horizon. Using $\partial_udA_\epsilon=\theta_udA_\epsilon$, we also have
\begin{equation}
  \partial_u^2dA_\epsilon
  =(\partial_u\theta_u+\theta_u^2)dA_\epsilon,
  \qquad
  \partial_u\delta dA_\epsilon
  =\delta\theta_u\,dA_\epsilon+\theta_u\,\delta dA_\epsilon\,.
\label{eq:area_element_derivatives}
\end{equation}
Substituting Eqs.~\eqref{eq:theta_u}, \eqref{eq:theta_relations}, and~\eqref{eq:area_element_derivatives} into Eq.~\eqref{eq:apparent_horizon_area_expansion_second_order1} gives
\begin{equation}
\begin{aligned}
  dA_{\mathcal T}(s;U,v,x)
  &=dA_\epsilon+s\left[
  \delta dA_\epsilon-v\delta\theta_vdA_\epsilon
  -vD_i(D^i\delta U+\beta^i\delta U)dA_\epsilon\right]\\
  &\quad+s^2\Biggl\{
  \frac12\delta^2dA_\epsilon-v\delta\theta_v\delta dA_\epsilon
  -\frac v2\delta^2\theta_vdA_\epsilon
  -vD_i(D^i\delta U+\beta^i\delta U)\delta dA_\epsilon\\
  &\quad+\frac12\left(1-\frac\lambda\mu\right)v^2
  \left[\delta\theta_v+D_i(D^i\delta U+\beta^i\delta U)\right]^2dA_\epsilon
  -\frac v2(D\delta U)^2\theta_udA_\epsilon\\
  &\quad-\frac v2D_i\Bigl[
  D^i\delta^2U+\beta^i\delta^2U
  +2(\delta\gamma^{ij}D_j\delta U+\delta\beta^i\delta U)
  +2\delta U(\partial_u\gamma^{ij}D_j\delta U
  +\delta U\partial_u\beta^i)\Bigr]dA_\epsilon\\
  &\quad+\Bigl[
  -\frac v2(D^i\delta U+\beta^i\delta U)
  D_i(\gamma^{jk}\delta\gamma_{jk})
  -v\delta U(D^i\delta U+\beta^i\delta U)D_i\theta_u\\
  &\quad
  +\delta U(\delta\theta_u-v\partial_u\delta\theta_v)
  +\frac12(\delta U)^2
  (\partial_u\theta_u-v\partial_u^2\theta_v)
  \Bigr]dA_\epsilon
  \Biggr\}+\mathcal O(s^3)\,.
\end{aligned}
\label{eq:apparent_horizon_area_expansion_second_order2}
\end{equation}
Similarly to the above calculation, we again suppress the explicit dependence of all quantities on the coordinates. Considering the result obtained in the first relation in Eq.~\eqref{eq:apparent_horizon_condition}, we get
\begin{equation}
\begin{aligned}
  dA_{\mathcal T}(s;U,v,x)
  &=dA_\epsilon+s\delta[(1-v\theta_v)dA_\epsilon]
  +\frac{s^2}{2}\delta^2[(1-v\theta_v)dA_\epsilon]
  -svD_i(D^i\delta U+\beta^i\delta U)dA_\epsilon\\
  &\quad+\frac{s^2}{2}\Biggl\{
  \left(1-\frac\lambda\mu\right)v^2
  \left[\delta\theta_v
  +D_i(D^i\delta U+\beta^i\delta U)\right]^2
  -v(D\delta U)^2\theta_u\\
  &\quad
  +2\delta U(\delta\theta_u-v\partial_u\delta\theta_v)
  +(\delta U)^2(\partial_u\theta_u-v\partial_u^2\theta_v)
  -2v\delta U(D^i\delta U+\beta^i\delta U)D_i\theta_u
  \Biggr\}dA_\epsilon\\
  &\quad-\frac{s^2}{2}vD_i\Bigl[
  D^i\delta^2U+\beta^i\delta^2U
  +2(\delta\gamma^{ij}D_j\delta U+\delta\beta^i\delta U)\\
  &\quad
  +2\delta U(\partial_u\gamma^{ij}D_j\delta U
  +\delta U\partial_u\beta^i)
  +(D^i\delta U+\beta^i\delta U)\gamma^{jk}\delta\gamma_{jk}
  \Bigr]dA_\epsilon+\mathcal O(s^3)\,.
\end{aligned}
\label{eq:apparent_horizon_area_expansion_second_order}
\end{equation}

In the following, we expand the apparent-horizon area in the perturbation parameter,
\begin{equation}
  A_{\mathcal T}(s;v)=A_{\mathcal T}(v)
  +s\delta A_{\mathcal T}(v)
  +\frac{s^2}{2}\delta^2A_{\mathcal T}(v)+\mathcal O(s^3)\,,
\end{equation}
and retain only the second-order perturbative contribution. We then integrate both sides of the equality in Eq.~\eqref{eq:apparent_horizon_area_expansion_second_order} over the codimension-2 horizon cross section at fixed $v$. At leading order, the area of $\mathcal{T}(v)$ is the same as that of $\mathcal{C}(v)$. It is also easy to find that at first order we have
\begin{equation}
  \delta A_{\mathcal T}(v)
  =\delta\int_{\mathcal C}dA_\epsilon(1-v\theta_v)\,.
\end{equation}
Since the last second-order term on the right-hand side of Eq.~\eqref{eq:apparent_horizon_area_expansion_second_order} is a total derivative, its integral vanishes. Thus, by using the first equation of Eq.~\eqref{eq:theta_relations}, we obtain
\begin{equation}
\begin{aligned}
  \delta^2A_{\mathcal T}(v)
  &=\delta^2\int_{\mathcal C}dA_\epsilon(1-v\theta_v)
  +\int_{\mathcal C}dA_\epsilon\Biggl\{
  \left(1+\frac{\lambda v}{\theta_u}\right)(\theta_u\delta U)^2
  -v(D\delta U)^2\theta_u\\
  &\quad
  +2\delta U(\delta\theta_u-v\partial_u\delta\theta_v)
  +(\delta U)^2(\partial_u\theta_u-v\partial_u^2\theta_v)
  -2v\delta U(D^i\delta U+\beta^i\delta U)D_i\theta_u
  \Biggr\}\,.
\end{aligned}
\label{eq:apparent_horizon_area_final}
\end{equation}
The additional integral is non-vanishing in general, but vanishes for $\delta U=0$.

\section[Modified Canonical Energy]{Modified Canonical Energy}
\label{sec:modified_canonical_energy}
In this section, we give a brief review of the covariant phase space formalism and introduce a modified canonical energy that will be used in the second-order analysis in the following sections. We include the effect of varying the vector field $\xi^a$, identify the relevant constraint and boundary terms, and evaluate the corresponding fluxes through $\mathcal{H}^+$ and $\mathcal{I}^+$. This yields the relation between the modified canonical energy and the second-order variation of the entropy.

\subsection{Covariant Phase Space Formalism}

First, we briefly review the covariant phase space formalism following the strategy of~\cite{Iyer:1994ys,Lee:1990nz,Hollands:2012sf}. Consider a diffeomorphism-covariant theory in $n$ dimensions with Lagrangian $n$-form $\boldsymbol L(\phi)$ whose variation takes the standard form
\begin{equation}
  \delta \boldsymbol L(\phi)
  = \boldsymbol E(\phi)\delta\phi
  + d\,\boldsymbol\Theta(\phi,\delta\phi) \, .
  \label{eq:lagrangian_variation}
\end{equation}
Here $\phi$ denotes the full set of dynamical fields including the metric and matter fields, $\boldsymbol E(\phi)=0$ gives the equations of motion, and $\boldsymbol\Theta(\phi,\delta\phi)$ is the symplectic potential $(n-1)$-form, which is linear in $\delta\phi$. The associated symplectic current $(n-1)$-form is
\begin{equation} \boldsymbol\omega(\phi;\delta_1\phi,\delta_2\phi) = \delta_1\boldsymbol\Theta(\phi,\delta_2\phi) - \delta_2\boldsymbol\Theta(\phi,\delta_1\phi) \, . 
\end{equation} 
In this definition, the variations entering the definition of $\boldsymbol\omega$ are field-space variations acting on the dynamical fields $\phi$. In particular, when one of the variations is $\delta_2\phi=\mathcal L_\xi\phi$, the vector field $\xi^a$ is held fixed in evaluating the corresponding field-space variation. Below, when $\xi^a$ is also allowed to vary, we denote by $\delta$ the total variation and by $\delta_\phi$ the variation with respect to the dynamical fields at fixed $\xi^a$. The contribution arising explicitly from $\delta\xi^a$ is then treated separately. Now let $\xi^a$ be an arbitrary smooth vector field. Replacing $\delta\phi$ in Eq.~\eqref{eq:lagrangian_variation} by $\mathcal{L}_\xi\phi$, namely the variation induced by the diffeomorphism generated by $\xi^a$, we can get
\begin{equation}
  \mathcal{L}_\xi\boldsymbol L(\phi)
  = d[\xi\cdot\boldsymbol L(\phi)]
  = \boldsymbol E(\phi)\mathcal{L}_\xi\phi
  + d\boldsymbol\Theta(\phi,\mathcal{L}_\xi\phi) \, .
  \label{eq:diffeomorphism_covariant_lagrangian_variation}
\end{equation}
The first equality follows from the Cartan magic formula
\begin{equation}
  \mathcal{L}_\xi\boldsymbol L(\phi)
  = d[\xi\cdot\boldsymbol L(\phi)] + \xi\cdot d\boldsymbol L(\phi) \, ,
\label{CK}
\end{equation}
where $``\cdot"$ denotes the contraction of a vector field with the first index of a differential form. When the fields are on-shell, i.e., $\boldsymbol E(\phi)=0$, Eq.~\eqref{eq:diffeomorphism_covariant_lagrangian_variation} implies that there exists a  Noether current $(n-1)$-form
\begin{equation}
  \boldsymbol J_\xi(\phi)
  = \boldsymbol\Theta(\phi,\mathcal{L}_\xi\phi) - \xi\cdot\boldsymbol L(\phi) \, ,
  \label{eq:noether_current_definition}
\end{equation}
which is closed. The Noether current also satisfies an off-shell identity and can equivalently be written as~\cite{Iyer:1995kg}
\begin{equation}
  \boldsymbol J_\xi(\phi) = d\boldsymbol Q_\xi(\phi) + \xi^a\boldsymbol C_a(\phi) \, ,
  \label{eq:noether_current_decomposition}
\end{equation}
where $\boldsymbol Q_\xi(\phi)$ is the Noether-charge $(n-2)$-form, and $\boldsymbol C_a(\phi)$ is a dual vector-valued $(n-1)$-form that vanishes on shell~\cite{Hollands:2024vbe}
\begin{eqnarray}
  \boldsymbol C_a(\phi)
  &=&\boldsymbol\epsilon_c^{(n)}
  \Bigg{[}
    2(E_G)^{c}{}_{a}
    - \sum_A A^{d_1\cdots d_k}{}_{b_1\cdots a \cdots b_l}\,
    (E_M)_{d_1\cdots d_k}{}^{b_1\cdots c \cdots b_l}
    \nonumber\\
    &&+ \sum_A A^{d_1\cdots c \cdots d_k}{}_{b_1\cdots b_l}\,
    (E_M)_{d_1\cdots a \cdots d_k}{}^{b_1\cdots b_l}
  \Bigg{]} \, .
  \label{eq:full_Ca_expression}
\end{eqnarray}
Here, $\boldsymbol\epsilon_c^{(n)}=\epsilon_{c a_1\cdots a_{n-1}}$ is the spacetime volume form, $(E_G)_{ab}=0$ are the equations of motion for $g_{ab}$, and $(E_M)_{a_1\cdots a_k}{}^{b_1\cdots b_l}=0$ are the equations of motion for the matter field $A^{a_1\cdots a_k}{}_{b_1\cdots b_l}$.

In the following, we consider a variation of the Noether current $\boldsymbol J_\xi(\phi)$ in which both the dynamical fields $\phi$ and the  vector field $\xi^a$ are varied. Varying Eq.~\eqref{eq:noether_current_definition}, we obtain
\begin{equation}
  \begin{aligned}
    \delta\boldsymbol J_\xi(\phi)
    &= \delta_\phi\boldsymbol\Theta(\phi,\mathcal{L}_\xi\phi)
    + \boldsymbol\Theta(\phi,\mathcal{L}_{\delta\xi}\phi)
    - \delta\xi\cdot\boldsymbol L(\phi) - \xi\cdot\delta\boldsymbol L(\phi) \\
    &= \boldsymbol J_{\delta\xi}(\phi)
    -\xi\cdot[\boldsymbol E(\phi)\delta\phi]
    + \boldsymbol\omega(\phi;\delta\phi,\mathcal{L}_\xi\phi)
    + d[\xi\cdot\boldsymbol\Theta(\phi,\delta\phi)] \, ,
  \end{aligned}
  \label{eq:noether_current_variation}
\end{equation}
where
\begin{equation}
\delta_\phi\boldsymbol\Theta(\phi,\mathcal{L}_\xi\phi)
  = \delta\boldsymbol\Theta(\phi,\mathcal{L}_\xi\phi) -\boldsymbol\Theta(\phi,\mathcal{L}_{\delta\xi}\phi) \, .
\end{equation}
Substituting the variation of Eq.~\eqref{eq:noether_current_decomposition} into Eq.~\eqref{eq:noether_current_variation}, we obtain the so-called ``fundamental identity"~\cite{Hollands:2012sf}
\begin{equation}
  \boldsymbol\omega(\phi;\delta\phi,\mathcal{L}_\xi\phi)
  = \xi\cdot[\boldsymbol E(\phi)\delta\phi]
  + \xi^a\delta\boldsymbol C_a(\phi)
  + d[\delta_\phi\boldsymbol Q_\xi(\phi) - \xi\cdot\boldsymbol\Theta(\phi,\delta\phi)] \, ,
  \label{eq:varied_symplectic_current_identity}
\end{equation}
where
\begin{equation}
  \delta_{\phi}\boldsymbol Q_\xi(\phi)
  = \delta\boldsymbol Q_\xi(\phi)
 - \boldsymbol Q_{\delta\xi}(\phi) \, .
\end{equation}
In the following, we take $\xi^a$ to be the Killing vector field of the stationary background, i.e., $\mathcal{L}_\xi\phi=0$. Using this condition and varying Eq.~\eqref{eq:varied_symplectic_current_identity} once more, we obtain
\begin{equation}
  \begin{aligned}
    \boldsymbol\omega(\phi;\delta\phi,\mathcal{L}_\xi\delta\phi)
    + \boldsymbol\omega(\phi;\delta\phi,\mathcal{L}_{\delta\xi}\phi)
    =& \xi\cdot[\delta\boldsymbol E(\phi)\delta\phi]
    + \xi\cdot[\boldsymbol E(\phi)\delta^2\phi]
    + \xi^a\delta^2\boldsymbol C_a(\phi)
    + d[\delta_{\phi}^2\boldsymbol Q_\xi(\phi) - \xi\cdot\delta\boldsymbol\Theta(\phi,
    \delta\phi)] \\
    &+ \delta\xi\cdot[\boldsymbol E(\phi)\delta\phi] 
    + \delta\xi^a\delta\boldsymbol C_a(\phi)
    + d[\delta_\phi\boldsymbol Q_{\delta\xi}(\phi) - \delta\xi\cdot\boldsymbol\Theta(\phi,\delta\phi)] \, .
  \end{aligned}
\end{equation}
Tracing the dependence on $\delta\xi^a$ shows that the first term on the left-hand side corresponds to the first line on the right-hand side, whereas the second term corresponds to the second line, all of whose terms are linear in $\delta\xi^a$, therefore,  we obtain
\begin{equation}
\boldsymbol\omega(\phi;\delta\phi,\mathcal{L}_\xi\delta\phi)
  = \xi\cdot[\delta\boldsymbol E(\phi)\delta\phi]
  + \xi\cdot[\boldsymbol E(\phi)\delta^2\phi]
  + \xi^a\delta^2\boldsymbol C_a(\phi)
  + d[\delta^2_{\phi}\boldsymbol Q_\xi(\phi) - \xi\cdot\delta\boldsymbol\Theta(\phi,
  \delta\phi)]\, .
\label{eq:second_varied_symplectic_current_identity}
\end{equation}
So far, we have gathered enough formulas on covariant phase space formalism which is crucial to study the canonical energy for the perturbation fields.

When external matter fields are completely absent, taking a variation of Eq.~\eqref{eq:lagrangian_variation} shows that $d\bm{\omega}(\phi,\delta\phi,\mathcal{L}_\xi\delta\phi)=0$ provided the perturbations satisfy the linearized equations of motion and $\xi^a$ is fixed. In this case, the canonical energy of $\delta\phi$ can be defined as the integral of $\bm{\omega}(\phi,\delta\phi,\mathcal{L}_\xi\delta\phi)$ over a Cauchy surface~\cite{Hollands:2012sf}. In the next section, we introduce a modified canonical energy which incorporates external matter fields.

\subsection{Construction of the Modified Canonical Energy}

\begin{figure}[H]
    \centering
\begin{tikzpicture}[scale=1.2,line width=1pt]

\coordinate (B) at (-0.1,0);

\coordinate (L1) at (0,0);
\coordinate (T1) at (2,2);
\coordinate (R1) at (4,0);
\coordinate (D1) at (2,-2);
\coordinate (H1) at (0.5,0.5);
\coordinate (I1) at (3.5,0.5);
\coordinate (H2) at (1.5,1.5);
\coordinate (I2) at (2.5,1.5);

\draw (L1)--(T1)--(R1)--(D1)--cycle;

\node at (0.5,-1) {$\mathcal{H}^-$};


\node[right] at (3.2,-1) {$\mathcal{I}^-$};
\node[right] at (4.1,0)
{$S^{n-2}_{\infty}$};


\node[below] at (2.0,0.5)
{$\Sigma(t_1)$};


\coordinate (Bt) at (1.0,1.0);
\coordinate (Ct) at (3.0,1.0);

\fill[gray!15]
(H1)--(H2)--(I2)--(I1)--cycle;

\draw[gray!50!black,very thick]
(H2)--(I2);
\draw[gray!50!black,line width=2pt]
(H1)--(H2);
\draw[gray!50!black,line width=2pt]
(I2)--(I1);
\draw[gray!50!black,very thick]
(H1)--(I1);

\node[left] at (B) {$\mathcal{B}$};
\node[above] at (2.0,1.0)
{$\Sigma(t_2)$};

\node[above] at (0.6,0.9)
{$\mathcal{H}^+_{12}$};
\node[above] at (3.3,0.9)
{$\mathcal{I}^+_{12}$};

\end{tikzpicture}
 \caption{A schematic diagram of the compact integration region and its boundary in Eq.~\eqref{eq:stokes_modified_symplectic_current}. The compact region is represented by the shaded area, whose boundary consists of two spacelike slices, $\Sigma(t_1)$ and $\Sigma(t_2)$ with $t_2>t_1$, together with the corresponding segments of the future event horizon, $\mathcal{H}_{12}^+$, and future null infinity, $\mathcal{I}_{12}^{+}$. $\mathcal{B}$ denotes the bifurcation surface.}
\label{fig:placeholder}
\end{figure}
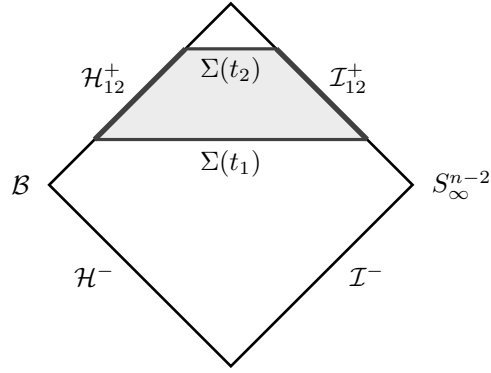

From this point on, we restrict attention to the case in which matter fields are merely absent in the background, and $\phi$ is identified with the metric field $g$ only. For a rotating background, the horizon-generating Killing field $\xi^a$ should be distinguished from the asymptotically timelike Killing field $t^a$:
\begin{equation}
\xi^a=t^a+\sum_I\Omega_I\psi_I^a\, .
\label{eq:horizon_and_stationary_generators}
\end{equation}
Here $I$ labels the independent rotational symmetries of the background, $\psi_I^a$ are the associated axial Killing fields, and $\Omega_I$ are the corresponding horizon angular velocities. The canonical energy defined below is constructed using $\xi^a$~\cite{Hollands:2012sf}. In order to define  a conserved canonical energy, we introduce a $(n-1)$-form $\bm{\omega}'(g;\delta g,\mathcal{L}_{\xi}\delta g)$ as follows
\begin{equation}
  \begin{aligned}
    \boldsymbol\omega'(g;\delta g,\mathcal{L}_\xi\delta g)
    &= \boldsymbol\omega(g;\delta g,\mathcal{L}_\xi\delta g)
    - \xi \cdot [\delta\boldsymbol E(g)\delta g]
    - \xi \cdot [\boldsymbol E(g)\delta^2 g]
    - \xi^a\delta^2\boldsymbol C_a(g) \, ,
  \end{aligned}
\label{eq:modified_symplectic_current_definition}
\end{equation}
where $\xi^a$ is the Killing vector field of the stationary background and
\begin{equation}
  d\boldsymbol\omega'(g;\delta g,\mathcal{L}_\xi\delta g)=0
\end{equation}
by considering Eq.~\eqref{eq:second_varied_symplectic_current_identity}. We refer to this $(n-1)$-form as the ``canonical energy form". The canonical energy of the perturbation on a Cauchy surface $\Sigma(t)$ is defined as
\begin{equation}
\mathcal{E}[\Sigma(t)] = \int_{\Sigma(t)} \bm{\omega}'(g;\delta g,\mathcal{L}_\xi\delta g)\, .
\label{new_canonical_energy}
\end{equation}
Here, the canonical-energy $(n-1)$-form is defined differently from the one in~\cite{Hollands:2012sf} to include the contribution associated with the external matter fields \footnote{According to the discussion in following sections, it is not hard to find that $$\mathcal{E}[\Sigma(t)]= \int_{\Sigma(t)}\bm{\omega} (g;\delta g\, ,\mathcal{L}_{\xi}\delta g)+ \int_{\Sigma(t)} \bm{\epsilon}^{(n-1)}\delta^2 T_{ab}\xi^a \nu^b\, ,$$ where $\nu^a$ is the unit normal vector field of the Cauchy surface $\Sigma(t)$, $\bm{\epsilon}^{(n-1)}$ is the volume element of $\Sigma(t)$, and $T_{ab}$ denotes the energy-momentum tensor of the external matter field whose first-order variation, i.e., $\delta T_{ab}$, is assumed to vanish (see the argument below Eq.(\ref{eq:deltaT_integral_zero}) or similar discussion in the appendix of ~\cite{Iyer:1994ys}). Obviously, the canonical energy includes the contribution from the pure gravitational perturbation as well as the one from the matter field. The latter is the same as the canonical energy of matter introduced in the appendix of~\cite{Iyer:1994ys}, where the spacetime metric is fixed as a background. Furthermore, $\boldsymbol{\omega}'(g;\delta_1 g,\mathcal{L}_\xi\delta_2 g)$ can be put into an exact form, i.e.,  $$\boldsymbol{\omega}'(g;\delta_1 g,\mathcal{L}_\xi\delta_2 g)=d[\delta_2\delta_1\boldsymbol{Q}_{\xi}(g)-\xi\cdot\delta_2\boldsymbol{\Theta}(g,\delta_1 g)]\, .$$ Using this expression, it is easy to find that the canonical energy defined here satisfies the Propositions $2$ and $3$ in~\cite{Hollands:2012sf}.}. This definition reduces to the one used in~\cite{Hollands:2012sf} only when external matter fields are completely absent. Therefore, by Stokes' theorem, in a compact region bounded by the Cauchy surfaces $\Sigma(t_2)$ and $\Sigma(t_1)$ with $t_2>t_1$, $\mathcal{H}^+_{12}$, and $\mathcal{I}^+_{12}$, we have
\begin{equation}
  \int_{\Sigma(t_1)}\boldsymbol\omega'
  = \int_{\Sigma(t_2)}\boldsymbol\omega'
  + \int_{\mathcal{H}^+_{12}}\boldsymbol\omega'
  + \int_{\mathcal{I}^+_{12}}\hat{\boldsymbol\omega}' \, .
  \label{eq:stokes_modified_symplectic_current}
\end{equation}
The hypersurfaces $\Sigma(t_2)$, $\Sigma(t_1)$, $\mathcal{H}_{12}^+$, and $\mathcal{I}_{12}^{+}$ in the above formula are illustrated schematically in Fig.~\ref{fig:placeholder}. \footnote{Here we have omitted some technical details, such as the existence of null infinity~\cite{Hollands:2003xp}.}

Here we have suppressed the dependence of $\boldsymbol\omega'$ on its arguments. The signs in this formula are determined by the boundary orientation. A hatted quantity denotes the corresponding unhatted quantity expressed in the unphysical spacetime and evaluated at conformal infinity $\mathcal{I}^+$. The region $\mathcal{H}^+_{12}$ denotes the portion of $\mathcal{H}^+$ bounded by the two cross sections specified by the affine parameters $v_1$ and $v_2$. Similarly, $\mathcal{I}^+_{12}$ denotes the portion of $\mathcal{I}^+$ bounded by the two cross sections specified by the affine parameters $ \hat{u}_1$ and $\hat{u}_2$. It follows that we can calculate the change of the canonical energy $\mathcal{E}[\Sigma(t)]$ between the two Cauchy surfaces $\Sigma(t_2)$ and $\Sigma(t_1)$ as
\begin{equation}
  {\mathcal{E}}[\Sigma(t_2)] - {\mathcal{E}}[\Sigma(t_1)]
  = - \int_{\mathcal{H}^+_{12}}\boldsymbol\omega' - \int_{\mathcal{I}^+_{12}}\hat{\boldsymbol\omega}' \, .
\label{eq:modified_canonical_energy_change}
\end{equation}
In order to calculate the contribution of $\boldsymbol\omega'(g;\delta g,\mathcal{L}_\xi\delta g)$ on $\mathcal{I}^+_{12}$, we need to specify the geometric structure near $\mathcal{I}^+_{12}$. For this purpose, we choose the Bondi gauge for the physical background metric $g_{ab}$~\cite{Hollands:2003xp,Hollands:2012sf}. The corresponding unphysical background metric $\hat g_{ab}=\Omega^2 g_{ab}$, where the conformal factor is $\Omega=1/r$, then admits, in a neighborhood of $\mathcal{I}^+$, the asymptotic form
\begin{equation}
  \hat g_{ab}
  = (d\Omega)_a (d\hat u)_b
  + (d\Omega)_b (d\hat u)_a
  + \hat\gamma_{ij} (dx^i)_a (dx^j)_b
  + \mathcal{O}(\Omega) \, .
  \label{eq:bondi_gauge_unphysical_metric}
\end{equation}
Here $\hat u$ is a future directed affine parameter on the null geodesic generators of $\mathcal{I}^+$. On $\mathcal{I}^+$, the generator is given by $\hat n^a\equiv(\partial/\partial\hat u)^a$, while $\hat\gamma_{ab}$ denotes the unit round metric on the $S^{n-2}$ cross sections of $\mathcal{I}^+$ and obeys $\hat\gamma_{ab}\hat n^a=0$ and $\hat\gamma_{ab}(\partial/\partial\Omega)^a=0$. It is the asymptotically timelike Killing field $t^a$ whose extension $\hat t^a$ is parallel to $\hat n^a$:
\begin{equation}
  \hat t^a=(\hat t^b\hat\nabla_b\hat u)\hat n^a\, ,
  \qquad
  \hat\xi^a=\hat t^a+\sum_I\Omega_I\hat\psi_I^a\, .
\label{eq:timelike_killing_field_at_null_infinity}
\end{equation}
The coefficient $\hat t^b\hat\nabla_b\hat u$ is positive and constant. We choose the cuts $\hat u=\mathrm{const.}$ of $\mathcal I^+$ such that the axial Killing fields $\hat\psi_I^a$ are tangent to them, so that $\hat\psi_I^a\hat\nabla_a\hat u=0$. For rotating backgrounds, we further restrict the null-infinity analysis to perturbations that preserve these axial symmetries,
\begin{equation}
\mathcal L_{\psi_I}\delta g_{ab}=0\quad\text{for every }I\, .
\label{eq:axisymmetric_perturbations}
\end{equation}
It then follows that $\mathcal L_\xi\delta g_{ab}=\mathcal L_t\delta g_{ab}$ and $\hat\xi^a\hat\nabla_a\hat u=\hat t^a\hat\nabla_a\hat u$, so that the flux at $\mathcal I^+$ takes the standard positive Bondi-news form~\cite{Hollands:2012sf}. This axisymmetry assumption is required only for the null-infinity flux analysis and not for the local horizon balance law.

In the following, we carry out the calculation by taking Einstein gravity as an example. We first evaluate the contribution from the horizon portion $\mathcal{H}^+_{12}$ in Eq.~\eqref{eq:modified_canonical_energy_change}. Using Eq.~(21) of~\cite{Hollands:2012sf}, we find that $\boldsymbol{\omega}(g,\delta g,\mathcal{L}_\xi \delta g)$ is given by
\begin{equation}
  \boldsymbol \omega(g;\delta g,\mathcal{L}_\xi \delta g)
  = \frac{1}{16\pi G} \, w^a \, \boldsymbol\epsilon_a^{(n)} \, ,
\end{equation}
where
\begin{equation}
  \begin{aligned}
    w^a
    &= P^{abcdef}\Bigl(
      \mathcal{L}_\xi \delta g_{bc}\,\nabla_d \delta g_{ef}
      - \delta g_{bc}\,\nabla_d \mathcal{L}_\xi \delta g_{ef}
    \Bigr) \, ,
  \end{aligned}
\end{equation}
and
\begin{equation}
  P^{abcdef}
  = g^{ae}g^{bf}g^{cd}
  - \frac{1}{2} g^{ad}g^{be}g^{cf}
  - \frac{1}{2} g^{ab}g^{cd}g^{ef}
  - \frac{1}{2} g^{bc}g^{ae}g^{df}
  + \frac{1}{2} g^{bc}g^{ad}g^{ef} \, .
\end{equation}
Using the stationarity condition on the background, one can show that
\begin{equation}
  \boldsymbol \omega(g;\delta g,\mathcal{L}_\xi \delta g)
  \overset{\mathcal{H}^+}{=}
  \frac{\kappa}{32\pi G} \, \boldsymbol{\epsilon}^{(n-1)}
  \left\{(\delta \gamma^{ab} + \gamma^{ab}\gamma^{cd}\delta \gamma_{cd})\partial_v(v\partial_v \delta \gamma_{ab}) - v\left[\partial_v\delta\gamma^{ab}\partial_v\delta\gamma_{ab} + 4(\delta\theta_v)^2\right]\right\} \, .
  \label{eq:omega_on_horizon}
\end{equation}
Here $\boldsymbol{\epsilon}^{(n-1)}$ is the volume $(n-1)$-form on $\mathcal{H}^+$. Then, the second and the third term in Eq.~\eqref{eq:modified_symplectic_current_definition} vanish when pulled back to $\mathcal{H}^+$. Finally, for the last term in Eq.~\eqref{eq:modified_symplectic_current_definition}, using~\eqref{eq:full_Ca_expression}, its second-order perturbative expansion simplifies to
\begin{equation}
  \xi^a\delta^2\boldsymbol C_a(g)
  \overset{\mathcal{H}^+}{=} -\kappa v[\delta \boldsymbol {\epsilon}^{(n-1)} 2 k^a k^b\delta T_{ab} + \boldsymbol {\epsilon}^{(n-1)}k^a k^b\delta^2T_{ab}] \, .
\label{eq:constraint_term_on_horizon}
\end{equation}
Here $\delta T_{ab}$ and $\delta^2 T_{ab}$ represent the first- and second-order stress-energy perturbations of the external infalling matter fields. They are determined by the perturbative gravitational equations of motion,
\begin{equation}
  2\delta (E_G)_{ab}=\delta T_{ab} \, , \qquad
  2\delta^2 (E_G)_{ab}=\delta^2 T_{ab} \, .
\end{equation}
In order to evaluate the horizon contribution appearing in Eq.~\eqref{eq:omega_on_horizon}, it is necessary to analyze the perturbations of the geometric quantities on the event horizon. First, we need to calculate the contribution from the shear tensor. The shear tensor is defined as
\begin{equation}
  \sigma_{ab}
  = \frac{1}{2} \mathcal{L}_k \gamma_{ab}
  - \frac{1}{n-2} \gamma_{ab} \theta_v \, .
\end{equation}
Expanding the Lie derivative along the $k$ direction and using $g_{ab} \overset{\mathcal{H}^+}{=}-2k_{(a}l_{b)} + \gamma_{ab}$, one obtains
\begin{equation}
  \sigma_{ab}
  = \frac{1}{2} \partial_v \gamma_{ab}
  - \frac{1}{n-2} \gamma_{ab} \theta_v \, .
\end{equation}
Here and in the following calculations, $\partial_v$ denotes $k^a\partial_a$, while $\partial_u$ denotes $l^a\partial_a$. After first-order perturbation and using the zeroth-order $\theta_v = 0$, one can find that
\begin{equation}
  \delta \sigma_{ab} = \frac{1}{2} \partial_v \delta \gamma_{ab} - \frac{1}{n-2} \gamma_{ab} \delta \theta_v \, , \qquad
  \delta \sigma^{ab} = - \frac{1}{2} \partial_v \delta \gamma^{ab} - \frac{1}{n-2} \gamma^{ab} \delta \theta_v \, .
\end{equation}
Using $\delta \theta_v = \frac{1}{2} \gamma^{ab} \partial_v \delta \gamma_{ab}$, the contraction of the first-order shear perturbation can be simplified as
\begin{equation}
  \delta \sigma_{ab} \, \delta \sigma^{ab}
  \overset{\mathcal{H}^+}{=}
  - \frac{1}{4} \partial_v \delta \gamma^{ab} \partial_v \delta \gamma_{ab}
  - \frac{( \delta \theta_v )^2}{ n-2 } \, .
\label{eq:first_order_shear_contraction}
\end{equation}
Second, we need to calculate the contribution from the twist tensor. The twist tensor is defined as
\begin{equation}
  \omega_{ab} = \gamma_a{}^c\gamma_b{}^d \nabla_{[c}k_{d]} \, ,
\end{equation}
where the projection operator $\gamma_a{}^b \overset{\mathcal{H}^+}{=} \delta_a{}^b + k_al^b + l_ak^b$. In the stationary case, we have
\begin{equation}
  \omega_{ab} = \gamma_a{}^c\gamma_b{}^d \partial_{[c} k_{d]} \overset{\mathcal{H}^+}{=} 0 \, , \qquad
  \omega^{ab} = g^{ac}g^{bd}\omega_{cd} \overset{\mathcal{H}^+}{=} 0 \, .
\end{equation}
The gauge conditions imply $\delta\gamma_a{}^b \overset{\mathcal{H}^+}{=}0$ and ensure that $k^a$ remains normal to $\mathcal{H}^+$. Hence,
\begin{equation}
  \delta\omega_{ab}
  = \gamma_a{}^c \gamma_b{}^d \delta\!\left(\partial_{[c} k_{d]}\right) \overset{\mathcal{H}^+}{=} 0 \, , \qquad
  \delta\omega^{ab} = g^{ac}g^{bd}\delta\omega_{cd} \overset{\mathcal{H}^+}{=} 0 \, .
\end{equation}
Finally, to obtain the relation between the relevant geometric quantities and the expansion, we introduce the Raychaudhuri equation for the null geodesic congruence generated by $k^a$~\cite{Hawking:1973qla,Kar:2006ms}:
\begin{equation}
  \frac{d\theta_v}{dv}
  = -\frac{1}{n-2}\,\theta_v^2
  - \sigma^{ab}\sigma_{ab}
  + \omega^{ab}\omega_{ab}
  - R_{ab}k^a k^b \, .
  \label{eq:raychaudhuri_horizon}
\end{equation}
Here $\sigma_{ab}$ and $\omega_{ab}$ are the shear and twist of the null congruence. Notice that $\omega_{ab}\overset{\mathcal{H}^+}{=}0$ since $k^a$ is hypersurface orthogonal on $\mathcal{H}^+$, and according to our calculation of the twist tensor, we also have $\delta\omega_{ab}\overset{\mathcal{H}^+}{=}0$. Moreover, $\theta_v$ and $\sigma_{ab}$ vanish in the background. Furthermore, according to the gauge conditions, the perturbation of the Einstein equation satisfies $\delta R_{ab} k^a k^b = 8\pi G\, \delta T'_{ab} k^a k^b$ and $\delta^2 R_{ab} k^a k^b = 8\pi G\, \delta^2 T'_{ab} k^a k^b$. Here $T'_{ab}$ denotes the total stress-energy tensor, including the contributions from all matter fields. Thus, the first-order perturbation and the second-order perturbation of the Raychaudhuri equation give
\begin{equation}
  \frac{d\delta\theta_v}{dv}
  \overset{\mathcal{H}^+}{=} - 8\pi G\, \delta T'_{ab} k^a k^b \, , \qquad
  \frac{d\delta^2\theta_v}{dv}
  \overset{\mathcal{H}^+}{=} \frac{1}{2} \partial_v \delta \gamma^{ab} \partial_v \delta \gamma_{ab} - 8\pi G \delta^2 T'_{ab} k^a k^b \, .
  \label{eq:first_second_order_raychaudhuri}
\end{equation}
Since we only consider contributions from external matter fields, $T'_{ab}$ is equivalent to $T_{ab}$ in the following analysis. Combining Eqs.~\eqref{eq:omega_on_horizon}, \eqref{eq:constraint_term_on_horizon}, and~\eqref{eq:first_order_shear_contraction}, then integrating over $\mathcal{H}^+_{12}$, we obtain the contribution of $\boldsymbol\omega'$ on $\mathcal{H}^+$.
\begin{equation}
  \begin{aligned}
\int_{\mathcal{H}^+_{12}}\boldsymbol\omega'
    =& \frac{1}{4\pi G}\int_{\mathcal H^+_{12}}\boldsymbol {\epsilon}^{(n-1)}(\xi^c\nabla_c v) \left[\delta\sigma^{ab}\delta\sigma_{ab} - \frac{n-3}{n-2}(\delta\theta_v)^2\right] \\
    &+ \frac{1}{32\pi G}\Delta \int_{\mathcal{C}}\boldsymbol {\epsilon}^{(n-2)}(\xi^c\nabla_c v)(\delta \gamma^{ab}\partial_v \delta \gamma_{ab} + 2\delta \theta_v \gamma^{ab}\delta \gamma_{ab}) \\
    &+ \int_{\mathcal{H}^+_{12}}(\xi^c\nabla_c v)[\delta \boldsymbol {\epsilon}^{(n-1)} 2 k^a k^b\delta T_{ab} + \boldsymbol {\epsilon}^{(n-1)}k^a k^b\delta^2 T_{ab}] \, .
  \end{aligned}
  \label{fluxathorizon1}
\end{equation}
The symbol $\Delta$ denotes the difference of a quantity evaluated between the two cross-sections $\mathcal{H}^+_1$ and $\mathcal{H}^+_2$.

We next evaluate the contribution from $\mathcal I^+_{12}$ to the change of $\mathcal E[\Sigma(t)]$. In even spacetime dimension, we adopt the asymptotic conditions of~\cite{Hollands:2012sf}, keeping $\Omega$ and the unit round metric $\hat\gamma_{ab}$ fixed. We define the rescaled inverse-metric perturbation
\begin{equation}
  \begin{aligned}
  \hat\tau^{ab}\equiv\Omega^{-\frac{n-2}{2}}\delta\hat g^{ab}\, ,\qquad
  \hat\tau_{ab}=-\Omega^{-\frac{n-2}{2}}\delta\hat g_{ab}\, .
  \end{aligned}
  \label{eq:rescaled_perturbation_null_infinity}
\end{equation}
This differs by an overall sign from the rescaled covariant perturbation in Eq.~(90) of~\cite{Hollands:2012sf}. It extends smoothly to $\mathcal I^+$, with $\hat g^{ab}\hat\tau_{ab}=\mathcal O(\Omega)$ and $\hat\tau_{ab}\hat n^a=\mathcal O(\Omega)$; its boundary value is understood below. Indices on $\hat\tau_{ab}$ are raised and lowered with $\hat g_{ab}$. The pullback of the symplectic current to $\mathcal I^+$ is related to the boundary potential by
\begin{equation}
  \hat{\boldsymbol\omega}
  (\hat g;\delta_1\hat g,\delta_2\hat g)
  \overset{\mathcal I^+}{=}
  \delta_1\hat{\boldsymbol\Theta}'
  (\hat g,\delta_2\hat g)
  -\delta_2\hat{\boldsymbol\Theta}'
  (\hat g,\delta_1\hat g)\, .
  \label{eq:symplectic_current_definition_null_infinity}
\end{equation}
In the convention of Eq.~\eqref{eq:rescaled_perturbation_null_infinity}, the boundary potential is~\cite{Wald:1999wa,Hollands:2012sf}
\begin{equation}
  \hat{\boldsymbol\Theta}'(\hat g,\delta\hat g)
  \overset{\mathcal I^+}{=}
  \frac{1}{32\pi G}\hat N^{ab}\hat\tau_{ab}
  \hat{\boldsymbol\epsilon}^{(n-1)}\, .
  \label{eq:theta_prime_null_infinity}
\end{equation}
Here $\hat N_{ab}$ is the Bondi news tensor, defined by
\begin{equation}
  \hat N_{ab}
  = \Omega^{-\frac{n-4}{2}}\left(\hat\gamma_a{}^{c}\hat\gamma_b{}^{d} - \frac{1}{n-2}\hat\gamma_{ab}\hat \gamma^{ef}\hat\gamma_e{}^{c}\hat\gamma_f{}^{d}\right)\left[\frac{2}{n-2}\hat R_{cd} - \frac{1}{(n-1)(n-2)}\hat R\,\hat g_{cd} \right] \, .
\label{eq:bondi_news_tensor_null_infinity}
\end{equation}
Here, $\hat{R}_{ab}$ and $\hat{R}$ are the Ricci tensor and Ricci scalar associated with the unphysical metric $\hat{g}_{ab}$, respectively. Projecting onto the spherical cross-sections $\hat{\mathcal C}$ with $\hat\gamma_a{}^b$, the linearized news satisfies~\cite{Hollands:2012sf}
\begin{equation}
  \delta\hat N_{ab}
  \overset{\mathcal I^+}{=}
  \hat\gamma_a{}^c\hat\gamma_b{}^d\mathcal L_{\hat n}\hat\tau_{cd}\, .
\end{equation}
Using this relation and Eq.~\eqref{eq:theta_prime_null_infinity} in Eq.~\eqref{eq:symplectic_current_definition_null_infinity}, with $\hat N_{ab}=0$ on the background, gives
\begin{equation}
  \begin{aligned}
  \hat{\boldsymbol\omega}(\hat g;\delta\hat g,\mathcal L_{\hat\xi}\delta\hat g) \overset{\mathcal I^+}{=}
  \frac{1}{32\pi G}\hat{\boldsymbol\epsilon}^{(n-1)}\hat\gamma^{ac}\hat\gamma^{bd} \Bigl[(\mathcal L_{\hat\xi}\hat\tau_{ab})(\mathcal L_{\hat n}\hat\tau_{cd})
  -\hat\tau_{ab}\mathcal L_{\hat\xi}\mathcal L_{\hat n}\hat\tau_{cd}\Bigr]\, .
  \end{aligned}
  \label{eq:null_infinity_symplectic_current_news}
\end{equation}
Here $\hat{\boldsymbol\epsilon}^{(n-1)}$ is the volume $(n-1)$-form on $\mathcal{I}^+$. Similarly, as on $\mathcal{H}^+$, the second and the third term in Eq.~\eqref{eq:modified_symplectic_current_definition} vanish when pulled back to $\mathcal{I}^+$. Finally, we consider the contribution from the last term of Eq.~\eqref{eq:modified_symplectic_current_definition}. We now explain why the corresponding integral over future null infinity does not contribute. In the present setup, the perturbation is produced by external matter fields falling into the black hole through the future event horizon $\mathcal{H}^+$. Consequently, the corresponding stress-energy perturbations are localized near the black hole and do not give rise to outgoing matter flux through future null infinity $\mathcal{I}^+$. Under the standard asymptotic flatness conditions, the matter stress tensor decays sufficiently rapidly toward null infinity, so that $\delta\hat T_{ab}\to0$ and $\delta^2\hat T_{ab}\to0$ at $\mathcal{I}^+$. As a result, the pullback of the last term of Eq.~\eqref{eq:modified_symplectic_current_definition} to $\mathcal I^+$ vanishes. Thus, the last term of Eq.~\eqref{eq:modified_symplectic_current_definition} contributes only on the horizon $\mathcal{H}^+$, while the flux through $\mathcal{I}^+$ is entirely determined by the radiative gravitational and matter degrees of freedom appearing in the symplectic current. According to Eq.~\eqref{eq:null_infinity_symplectic_current_news} and integrating over $\mathcal{I}^+_{12}$, we obtain the contribution of $\boldsymbol\omega'$ on $\mathcal{I}^+$.
\begin{equation}
  \begin{aligned}
\int_{\mathcal{I}^+_{12}}\hat{\boldsymbol\omega}'
  = \frac{1}{16\pi G}\int_{\mathcal{I}^+_{12}}\hat{\boldsymbol\epsilon}^{(n-1)}(\hat t^c\hat\nabla_c\hat u) \delta \hat N^{ab}\delta \hat N_{ab}
  - \frac{1}{32\pi G}\Delta\int_{\hat{\mathcal C}}\hat{\boldsymbol\epsilon}^{(n-2)}(\hat t^c\hat\nabla_c\hat u)\hat\tau^{ab}\delta\hat N_{ab} \, .
  \end{aligned}
  \label{fluxatinfinity}
\end{equation}
The symbol $\Delta$ denotes the difference of a quantity evaluated between the two cross-sections $\mathcal{I}^+_1$ and $\mathcal{I}^+_2$.

From the calculations above, we see that the change of $\mathcal{E}[\Sigma(t)]$ not only depends on the fluxes through $\mathcal{H}^+_{12}$ and $\mathcal{I}^+_{12}$, but also on two boundary terms. This motivates us to subtract these boundary terms and define a modified canonical energy
\begin{equation}
  \begin{aligned}
\mathcal{E}'[\Sigma(t)]
  \equiv\mathcal{E}[\Sigma(t)]
  + \frac{1}{32\pi G}\Bigg[\int_{\mathcal{C}}\boldsymbol {\epsilon}^{(n-2)}(\xi^c\nabla_c v)(\delta \gamma^{ab}\partial_v \delta \gamma_{ab} + 2\delta \theta_v \gamma^{ab}\delta \gamma_{ab})
  - \int_{\hat{\mathcal C}}\hat{\boldsymbol\epsilon}^{(n-2)}(\hat t^c\hat\nabla_c\hat u)\hat\tau^{ab}\delta\hat N_{ab}\Bigg] \, .
  \end{aligned}
  \label{eq:modified_canonical_energy_prime_definition}
\end{equation}
Consequently, the change of the modified canonical energy $\mathcal{E}'[\Sigma(t)]$ between two Cauchy surfaces $\Sigma(t_2)$ and $\Sigma(t_1)$ is given by
\begin{equation}
  \begin{aligned}
    {\mathcal{E}'}[\Sigma(t_2)] - {\mathcal{E}'}[\Sigma(t_1)]
    ={}& -\frac{1}{4\pi G}\int_{\mathcal H^+_{12}}\boldsymbol {\epsilon}^{(n-1)}(\xi^c\nabla_c v)\left[\delta\sigma^{ab}\delta\sigma_{ab} - \frac{n-3}{n-2}(\delta\theta_v)^2\right] \\
    &- \int_{\mathcal{H}^+_{12}}(\xi^c\nabla_c v)[\delta \boldsymbol {\epsilon}^{(n-1)} 2 k^a k^b\delta T_{ab} + \boldsymbol {\epsilon}^{(n-1)}k^a k^b\delta^2T_{ab}] \\
    &- \frac{1}{16\pi G}\int_{\mathcal{I}^+_{12}}\hat{\boldsymbol\epsilon}^{(n-1)}(\hat t^c\hat\nabla_c\hat u) \delta \hat N^{ab}\delta \hat N_{ab} \, .
\label{eq:canonical_energy_entropy_relation}
  \end{aligned}
\end{equation}
It follows that the change of the modified canonical energy $\mathcal{E}'[\Sigma(t)]$ is totally determined by the fluxes on the horizon and null infinity, which is in sharp contrast to the canonical energy $\mathcal{E}[\Sigma(t)]$.

\section[entropy]{Dynamical black hole entropy}
\label{sec:dynamical_entropy}
In this section, we combine the geometric results with the modified canonical energy to compare the entropy with the apparent-horizon area at second order. After evaluating the relevant boundary and flux terms, we use the Raychaudhuri equation and the shear contribution to simplify the entropy variation.

Our result shows that, at second order, the entropy remains proportional to the area of the apparent horizon under the null energy condition, whereas additional conditions are required to establish the same proportionality without the null energy condition.

\subsection{Derivation of the Balance Law}

In the following, we show that the two boundary terms in Eq.~\eqref{eq:modified_canonical_energy_prime_definition} can be canceled by the integrals of two exact forms, i.e.,
\begin{equation}
d\!\left[\xi \cdot \delta \boldsymbol\Theta(g,\delta g) - \xi \cdot \delta^2 \boldsymbol B_{\mathcal{H}^+}(g)\right]\, ,
\label{exactformhorizon}
\end{equation}
and
\begin{equation}
d\!\left[\hat\xi \cdot \delta \hat{\boldsymbol\Theta}(\hat g,\delta\hat g) - \hat\xi \cdot \delta^2 \hat{\boldsymbol B}_{\mathcal{I}^+}(\hat g)\right]
\label{exactforminfinity}
\end{equation}
on $\mathcal{H}^+_{12}$ and $\mathcal{I}^+_{12}$, respectively. Here $\boldsymbol B_{\mathcal{H}^+}(g)$ is defined by
\begin{equation}
  \underline{\boldsymbol\Theta}(g,\delta g)\overset{\mathcal{H}^+}{=}\delta\boldsymbol B_{\mathcal{H}^+}(g) \, 
  \label{eq:horizon_counterterm_definition}
\end{equation}
where $\underline{\boldsymbol\Theta}(g,\delta g)$ denotes the pullback of $\boldsymbol\Theta(g,\delta g)$ to the event horizon. This equation shows that, for a stationary background, there exists a quantity $\boldsymbol B_{\mathcal{H}^+}(g)$ on $\mathcal{H}^+$ such that $\boldsymbol\Theta(g,\delta g)$ is exact in field space when pulled back to $\mathcal{H}^+$, with $\boldsymbol B_{\mathcal{H}^+}(g)\overset{\mathcal{H}^+}{=}0$ on the stationary background (see Sec.~4.7 of~\cite{Hollands:2024vbe}). Moreover, $\hat{\boldsymbol B}_{\mathcal{I}^+}(\hat g)$ and $\hat{\boldsymbol\Theta}^{\prime}(\hat g,\delta\hat g)$ are given by~\cite{Wald:1999wa}
\begin{equation}
  \delta \hat{\boldsymbol B}_{\mathcal{I}^+}(\hat g)
  \overset{\mathcal{I}^+}{=}
  \hat{\boldsymbol\Theta}(\hat g,\delta\hat g)
  - \hat{\boldsymbol\Theta}'(\hat g,\delta\hat g) \, .
\label{eq:null_infinity_counterterm_definition}
\end{equation}
Here $\hat{\boldsymbol\Theta}'(\hat g,\delta\hat g)$ denotes the symplectic potential associated with the pullback of $\hat{\boldsymbol\omega}(\hat g,\delta\hat g,\mathcal{L}_{\hat\xi}\delta\hat g)$ to $\mathcal{I}^+$.

We first evaluate the contribution from the term Eq.~\eqref{exactformhorizon}. Here the symplectic potential is given by~\cite{Hollands:2024vbe}
\begin{equation}
  \boldsymbol\Theta(g,\delta g)
  = \frac{1}{16\pi G} \, \boldsymbol\epsilon_a^{(n)} g^{ab} g^{cd}
  \left(\nabla_c \delta g_{bd} - \nabla_b \delta g_{cd}\right) ,
\end{equation}
and
\begin{equation}
  \boldsymbol B_{\mathcal{H}^+}(g)
  = \frac{1}{16\pi G} \, l^a \boldsymbol\epsilon_{a}^{(n)} g^{bc} \mathcal{L}_k g_{bc}
  = \frac{1}{8\pi G} \, \boldsymbol\epsilon^{(n-1)}\, \theta_v \, .
  \label{eq:BH_boundary_term_definition}
\end{equation}
Using the gauge conditions together with the stationary condition on the background, one finds  that
\begin{equation}
  \delta\boldsymbol\Theta(g,\delta g)
  = \frac{1}{16\pi G} \, \boldsymbol\epsilon^{(n-1)}(2\delta^2 \theta_v + \gamma^{ab} \delta\gamma_{ab} \delta\theta_v - \delta^2\gamma^{ab} \partial_v\gamma_{ab} - \frac{1}{2}\delta\gamma^{ab} \partial_v\delta\gamma_{ab}) \, ,
  \label{eq:delta_theta_on_horizon}
\end{equation}
and
\begin{equation}
  \delta^2\boldsymbol B_{\mathcal{H}^+}(g)
  = \frac{1}{8\pi G} \, \boldsymbol\epsilon^{(n-1)}(\delta^2\theta_v + \gamma^{ab} \delta\gamma_{ab} \delta\theta_v) \, .
  \label{eq:delta2_BH_on_horizon}
\end{equation}
Notice that for an $(n-1)$-form ${\boldsymbol p}$, the pullback of $\xi \cdot d{\boldsymbol p}$ to the horizon vanishes. Using Eq.~\eqref{CK} to simplify the expression, we obtain the following result.
\begin{equation}
  \begin{aligned}
    d\!\left[\xi \cdot \delta \boldsymbol\Theta(g,\delta g) - \xi \cdot \delta^2 \boldsymbol B_{\mathcal{H}^+}(g)\right]
    \overset{\mathcal{H}^+}{=}& - \frac{\kappa}{32\pi G}\boldsymbol {\epsilon}^{(n-1)}(2\gamma^{ab}\delta \gamma_{ab}\delta \theta_v + \delta \gamma^{ab}\partial_v \delta \gamma_{ab}) \\
    &- \frac{\kappa}{32\pi G}\boldsymbol {\epsilon}^{(n-1)} v\partial_v(2\gamma^{ab}\delta \gamma_{ab}\delta \theta_v + 2\delta^2 \gamma^{ab}\partial_v \gamma_{ab} + \delta \gamma^{ab}\partial_v \delta \gamma_{ab}) \, .
  \end{aligned}
  \label{eq:boundary_term_on_horizon}
\end{equation}
Integrating this expression over $\mathcal{H}^+_{12}$ gives the corresponding boundary contribution on the horizon.
\begin{equation}
  \int_{\mathcal{H}^+_{12}}d\!\left[\xi \cdot \delta \boldsymbol\Theta(g,\delta g) - \xi \cdot \delta^2 \boldsymbol B_{\mathcal{H}^+}(g)\right]
  = -\frac{1}{32\pi G}\Delta \int_{\mathcal{C}}\boldsymbol {\epsilon}^{(n-2)}(\xi^c\nabla_c v)(\delta \gamma^{ab}\partial_v \delta \gamma_{ab} + 2\delta \theta_v \gamma^{ab}\delta \gamma_{ab}) \, .
  \label{eq:horizon_boundary_integral}
\end{equation}
We then calculate the contribution from the term in Eq.~\eqref{exactforminfinity}. Substituting Eq.~\eqref{eq:theta_prime_null_infinity} into this term and using Eq.~\eqref{eq:null_infinity_counterterm_definition}, one obtains
\begin{equation}
  d\!\left[\hat{\xi} \cdot \delta \hat{\boldsymbol\Theta}(\hat{g},\delta \hat{g}) - \hat{\xi} \cdot \delta^2 \hat{\boldsymbol B}_{\mathcal{I}^+}(\hat{g})\right]
  \overset{\mathcal{I}^+}{=}\frac{1}{32\pi G}\hat{\boldsymbol\epsilon}^{(n-1)}\hat \xi^c \partial_c(\hat\tau^{ab}\delta\hat N_{ab}) \, .
  \label{eq:boundary_term_null_infinity}
\end{equation}
Integrating this expression over $\mathcal{I}^+_{12}$ gives the corresponding boundary contribution at future null infinity.
\begin{equation}
  \int_{\mathcal{I}^+_{12}}d\!\left[\hat{\xi} \cdot \delta \hat{\boldsymbol\Theta}(\hat{g},\delta \hat{g}) - \hat{\xi} \cdot \delta^2 \hat{\boldsymbol B}_{\mathcal{I}^+}(\hat{g})\right]
  = \frac{1}{32\pi G}\Delta\int_{\hat{\mathcal C}}\hat{\boldsymbol\epsilon}^{(n-2)}(\hat t^c\hat\nabla_c\hat u)\hat\tau^{ab}\delta\hat N_{ab} \, .
  \label{eq:null_infinity_boundary_integral}
\end{equation}
Comparing Eqs.~\eqref{eq:horizon_boundary_integral} and \eqref{eq:null_infinity_boundary_integral} with Eq.~\eqref{eq:modified_canonical_energy_prime_definition}, we see that the boundary terms arising from difference of the last two terms in Eq.~\eqref{eq:modified_canonical_energy_prime_definition} between $\Sigma(t_1)$ and $\Sigma(t_2)$, except for opposite signs, are precisely given by the integrals of the exact forms \eqref{exactformhorizon} and \eqref{exactforminfinity} over $\mathcal{H}^+_{12}$ and $\mathcal{I}^+_{12}$, respectively.

Based on this and according to Eq.~\eqref{eq:modified_symplectic_current_definition}, we can construct the corresponding ``modified canonical energy $(n-1)$-forms" $\boldsymbol e_G$ on $\mathcal{H}^+$
\begin{equation}
  \boldsymbol e_G
  \overset{\mathcal{H}^+}{=} \boldsymbol \omega'(g;\delta g,\mathcal{L}_\xi \delta g)
  + d\!\left[\xi \cdot \delta \boldsymbol\Theta(g,\delta g) - \xi \cdot \delta^2 \boldsymbol B_{\mathcal{H}^+}(g)\right] \, ,
\label{eq:modified_canonical_energy_definition}
\end{equation}
and $\hat{\boldsymbol e}_G$ on $\mathcal{I}^+$
\begin{equation}
  \hat{\boldsymbol e}_G
  \overset{\mathcal{I}^+}{=} \hat{\boldsymbol\omega}'(\hat{g};\delta \hat{g},\mathcal{L}_{\hat{\xi}} \delta \hat{g})
  + d\!\left[\hat{\xi} \cdot \delta \hat{\boldsymbol\Theta}(\hat{g},\delta \hat{g}) - \hat{\xi} \cdot \delta^2 \hat{\boldsymbol B}_{\mathcal{I}^+}(\hat{g})\right]
   \, .
\end{equation}
So Eq.~\eqref{eq:canonical_energy_entropy_relation} can be written as
\begin{equation}
  \mathcal{E}'[\Sigma(t_2)]-\mathcal{E}'[\Sigma(t_1)]=-\int_{\mathcal{H}^+_{12}}\bm{e}_G -  \int_{\mathcal{I}^+_{12}}\hat{\bm{e}}_G\, .
\end{equation}
Considering the fact that the second and third terms in Eq.~\eqref{eq:modified_symplectic_current_definition} vanish on $\mathcal{H}^+_{12}$ and $\mathcal{I}^+_{12}$, and substituting the identity \eqref{eq:second_varied_symplectic_current_identity}, we find
\begin{equation}
  \boldsymbol e_G
   \overset{\mathcal{H}^+}{=}  d\!\left[\delta_{\phi}^2 \boldsymbol Q_\xi(g) - \xi \cdot \delta^2 \boldsymbol B_{\mathcal{H}^+}(g)\right] \, ,
\label{eq:modified_canonical_energy_definition1}
\end{equation}
and
\begin{equation}
  \hat{\boldsymbol e}_G
  \overset{\mathcal{I}^+}{=} d\!\left[\delta_{\phi}^2 \hat{\boldsymbol Q}_{\hat{\xi}}(\hat{g}) - \hat{\xi} \cdot \delta^2 \hat{\boldsymbol B}_{\mathcal{I}^+}(\hat{g})\right] \, .
\end{equation}
Based on the first law of black hole mechanics and some innovative thermodynamic consideration, the authors in~\cite{Hollands:2024vbe} introduce an entropy $(n-2)$-form
\begin{equation}
  \boldsymbol S \equiv \frac{2\pi}{\kappa_3}\left[\boldsymbol Q_\xi(g) - \xi \cdot \boldsymbol B_{\mathcal{H}^+}(g)\right] \, ,
\end{equation}
where $\kappa_3$ is the surface gravity defined as
\begin{equation}
2\kappa_3^2=-(\nabla_{[a}\xi_{b]})(\nabla^{[a}\xi^{b]})\, .
\end{equation}
The integral of $\bm{S}$ is nothing but the entropy of the system. Below, we will present the detailed relation between the modified canonical energy and the entropy.

To obtain the variation of entropy on $\mathcal{H}^+$, we need the entropy $(n-1)$-form on $\mathcal{H}^+$, namely
\begin{equation}
  d\boldsymbol S = 2\pi d\left[\frac{\boldsymbol Q_\xi(g)}{\kappa_3} - \frac{\xi \cdot \boldsymbol B_{\mathcal{H}^+}(g)}{\kappa_3}\right] \, .
  \label{eq:entropy_n_minus_one_form}
\end{equation}
Now, we will show that, under second-order perturbations, the entropy depends only on the variation of the fields and not on the variation of $\xi^a$ in \hyperref[sec:caseI]{Case I} below, but need not vanish for general perturbations. For general diffeomorphism covariant theories, the Noether charge $(n-2)$-form $\bm{Q}_{\xi}(g)$ has been thoroughly studied in~\cite{Iyer:1994ys}. However, for Einstein gravity, we simply have
\begin{equation}
\bm{Q}_{\xi}(g)= -\frac{1}{16\pi G} \bm{\epsilon}^{(n)}_{ab}\nabla^a\xi^b \, ,
\end{equation}
or
\begin{equation}
\bm{Q}_{\xi}(g)= -\frac{1}{16\pi G} \bm{\epsilon}^{(n-2)}(\bm{\epsilon}_{ab}\nabla^a\xi^b)\, ,
\end{equation}
where $\bm{\epsilon}_{ab}=k_al_b-l_ak_b$ is the so-called binormal when restricted on the event horizon. It is not hard to find ~\cite{Iyer:1994ys,Visser:2024pwz}
\begin{equation}
  \bm{\epsilon}^{ab}\nabla_a\xi_b
  = (k^al^b-k^bl^a)\nabla_a\xi_b=k^a\nabla_a(l^b\xi_b) - l^a\nabla_a (k^b\xi_b)= -\kappa_2 - \kappa_1=-2\kappa_3\, ,
\end{equation}
here we have used the fact that $l^a$ and $k^a$ commute to each other, and the definitions of the surface gravities $\kappa_1$, $\kappa_2$
\begin{equation}
  \nabla_a(\xi_b\xi^b)=-2\kappa_1\xi_a\, ,\qquad \xi^b\nabla_b\xi^a=\kappa_2\xi^a\,
\end{equation}
These can be equivalently expressed as
\begin{equation}
  \kappa_2=-k^a\nabla_a(l^b\xi_b)\, ,\qquad \kappa_1=l^a\nabla_a(k^b\xi_b)\, ,
\end{equation}
From the definition, it is easy to find $\kappa_3=(\kappa_1+\kappa_2)/2$, and also on the stationary background, the surface gravities satisfy $\kappa_2=\kappa_3=\kappa$. So, on the event horizon, we have
\begin{equation}
\bm{Q}_{\xi}(g)= \frac{\kappa_3}{8\pi G} \bm{\epsilon}^{(n-2)}\, .
\label{eq:noether_charge_horizon_slice}
\end{equation}
Moreover, on the horizon $\mathcal H^+$, the variation of surface gravity $\delta\kappa_2$ and $\delta\kappa_3$ satisfy (see Appendix~A of~\cite{Visser:2024pwz})
\begin{equation}
  \delta\kappa_2\overset{\mathcal H^+}{=}-k^a\nabla_a(l_b\delta\xi^b) \, , \qquad \delta\kappa_3\overset{\mathcal H^+}{=}l_{[a}k_{b]}\nabla^a\delta\xi^b \, .
  \label{eq:delta_kappa2_delta_kappa3_definitions}
\end{equation}
From the gauge condition, one obtains
\begin{equation}
\delta\xi^a\overset{\mathcal H^+}{=}\left(\int_0^v\delta\kappa_2\,dv'\right)k^a\, .
\label{deltaxiintegral}
\end{equation}
Next, we derive the full second-order variation of the two terms in Eq.~\eqref{eq:entropy_n_minus_one_form}. For the first term, we have
\begin{equation}
  \begin{aligned}
    \delta^2 \left[\frac{\boldsymbol Q_\xi(g)}{\kappa_3}\right]
    = \frac{1}{8\pi G}\delta_\phi^2\boldsymbol{\epsilon}^{(n-2)}
    = \frac{1}{\kappa}\delta^2_{\phi}\boldsymbol Q_\xi(g) \, .
  \end{aligned}
  \label{eq:second_variation_noether_charge_over_kappa}
\end{equation}
Similarly, for the second term, we have
\begin{equation}
  \delta^2\left[\frac{\xi\cdot\boldsymbol B_{\mathcal{H}^+}(g)}{\kappa_3}\right]
  \overset{\mathcal{H}^+}{=} \frac{1}{\kappa}\xi\cdot\delta^2_{\phi}\boldsymbol B_{\mathcal{H}^+}(g) + \frac{2}{\kappa}\left[\delta\xi\cdot\delta\boldsymbol B_{\mathcal H^+}(g) - \frac{\delta\kappa_3}{\kappa}\,\xi\cdot\delta\boldsymbol B_{\mathcal H^+}(g)
  \right] \, .
  \label{eq:second_variation_boundary_term_over_kappa}
\end{equation}
Combining Eqs.~\eqref{eq:entropy_n_minus_one_form}, \eqref{eq:second_variation_noether_charge_over_kappa}, and \eqref{eq:second_variation_boundary_term_over_kappa}, we then find that
\begin{eqnarray}
  \frac{\kappa}{2\pi}d\delta^2\boldsymbol S&\overset{\mathcal{H}^+}{=}&
   d\!\left[\delta_{\phi}^2 \boldsymbol Q_\xi(g) - \xi \cdot \delta^2 \boldsymbol B_{\mathcal{H}^+}(g)\right]
   - 2d\left[\delta\xi\cdot\delta\boldsymbol B_{\mathcal H^+}(g) - \frac{\delta\kappa_3}{\kappa}\xi\cdot\delta\boldsymbol B_{\mathcal H^+}(g)\right] \, .
\label{eq:second_order_entropy_form_horizon}
\end{eqnarray}
When $\delta\xi^a=0$ and only the variation of $g$ is involved, combining Eqs.~\eqref{eq:canonical_energy_entropy_relation}, \eqref{eq:modified_canonical_energy_definition}, and \eqref{eq:second_order_entropy_form_horizon}, then pulling the result back to $\mathcal{H}^+$, we obtain
\begin{equation}
\frac{\kappa}{2\pi}\Delta\delta_{\phi}^2S=\int_{\mathcal{H}^+_{12}}\boldsymbol e_G \, .
\label{eq:canonical_energy_dynamical_entropy_relation_old}
\end{equation}
We see that, under the second-order perturbations, the decrease of the modified canonical energy on $\mathcal{H}^+$ gives the variation of the entropy. We now consider the case $\delta\xi^a\neq0$. In other words, the variation involves not only the metric field $g$ but also  $\xi^a$. Therefore, the entropy formula should be corrected by the additional terms arising from the variation of $\xi^a$, and the corresponding relation becomes
\begin{equation}
\frac{\kappa}{2\pi} \, \Delta \delta^2 S=\int_{\mathcal{H}^+_{12}}\boldsymbol e_G
- \int_{\mathcal{H}^+_{12}}2d\left[\delta\xi\cdot\delta\boldsymbol B_{\mathcal H^+}(g) - \frac{\delta\kappa_3}{\kappa}\xi\cdot\delta\boldsymbol B_{\mathcal H^+}(g)\right] \, .
\label{eq:canonical_energy_dynamical_entropy_relation}
\end{equation}
Eqs.~\eqref{eq:canonical_energy_dynamical_entropy_relation_old} and ~\eqref{eq:canonical_energy_dynamical_entropy_relation} are the so-called ``balance law" that we expect to obtain.

\subsection{Relation to Apparent Horizon}

Now, we will derive the explicit expression for the entropy by evaluating the modified canonical energy and the additional terms. According to Eqs.~\eqref{eq:canonical_energy_entropy_relation}, \eqref{eq:modified_canonical_energy_definition}, and \eqref{eq:canonical_energy_dynamical_entropy_relation}, it follows that
\begin{equation}
  \begin{aligned}
\frac{\kappa}{2\pi} \, \Delta \delta^2 S     
    =& \kappa\int_{v_1}^{v_2} dv \int_{\mathcal{C}}(\delta dA_{\epsilon}\, 2v k^a k^b\delta T_{ab} + dA_{\epsilon}\, v k^a k^b\delta^2T_{ab})\\
    &+ \frac{\kappa}{4\pi G}\int_{v_1}^{v_2} dv \int_{\mathcal{C}} dA_{\epsilon} v \left[\delta\sigma^{ab}\delta\sigma_{ab} - \frac{n-3}{n-2}(\delta\theta_v)^2\right] \\
    &- \Delta\int_{\mathcal{C}}2
    \left[\delta\xi\cdot\delta\boldsymbol B_{\mathcal H^+}(g) - \frac{\delta\kappa_3}{\kappa}\xi\cdot\delta\boldsymbol B_{\mathcal H^+}(g)\right] \, .
  \end{aligned}
\end{equation}
Substituting the explicit form of $\delta \sigma_{ab} \delta \sigma^{ab}$ and Eqs.~\eqref{eq:first_order_shear_contraction}, \eqref{eq:first_second_order_raychaudhuri} and \eqref{eq:BH_boundary_term_definition}, we find
\begin{equation}
  \begin{aligned}
\frac{\kappa}{2\pi} \, \Delta \delta^2 S      
    =& - \frac{\kappa}{4\pi G} \int_{v_1}^{v_2} dv \int_{\mathcal{C}} \delta dA_{\epsilon}\, v  \frac{d \delta \theta_v}{dv} \\
    & - \frac{\kappa}{4\pi G} \int_{v_1}^{v_2} dv \int_{\mathcal{C}} dA_{\epsilon}\left[\frac{1}{2}v \frac{d \delta^2 \theta_v}{dv} +v (\delta \theta_v)^2\right] \\
    & + \frac{1}{4\pi G}\Delta\int_{\mathcal{C}}dA_{\epsilon}
    \left(v\delta\kappa_3 + \delta\xi\cdot l\right)\delta\theta_v \, .
  \end{aligned}
\end{equation}
Using the gauge condition, together with $\partial_v\delta dA_{\epsilon} = \delta\theta_v dA_{\epsilon}$ and integration by parts with respect to $v$, we obtain
\begin{equation}
  \begin{aligned}
\frac{\kappa}{2\pi} \, \Delta \delta^2 S  
    =& - \frac{\kappa}{4\pi G} \,\Delta \int_{\mathcal{C}} \delta dA_{\epsilon}\, v\, \delta \theta_v
    + \frac{\kappa}{4\pi G} \int_{v_1}^{v_2} dv \int_{\mathcal{C}} \delta dA_{\epsilon}\, \delta \theta_v \\
    & - \frac{\kappa}{8\pi G} \,\Delta \int_{\mathcal{C}} dA_{\epsilon}\, v\, \delta^2{\theta}_v
    + \frac{\kappa}{8\pi G} \int_{v_1}^{v_2} dv \int_{\mathcal{C}} dA_{\epsilon}\, \delta^2 \theta_v \\
    & + \frac{1}{4\pi G}\Delta\int_{\mathcal{C}}dA_{\epsilon}[\left(v\delta\kappa_3 - \delta h\right)\delta\theta_v] \, .
  \end{aligned}\label{Deltadelta2S}
\end{equation}
Integrating the first equation of Eq.~\eqref{eq:delta_kappa2_delta_kappa3_definitions} on $\mathcal H^+$, one obtains Eq.~\eqref{deltaxiintegral}. We then define
\begin{equation}
  \delta h \equiv \int_0^v \delta\kappa_2\,dv'\, . 
\end{equation} 
Together with $\theta_vdA_{\epsilon} = \partial_vdA_{\epsilon}$ and $\partial_v\delta dA_{\epsilon} = \delta\theta_vdA_{\epsilon}$, by exchanging the order of variation and integration, \eqref{Deltadelta2S} reduces to
\begin{equation}
\frac{\kappa}{2\pi} \, \Delta \delta^2 S  
  = \frac{\kappa}{8 \pi G} \Delta \delta^2 \int_{\mathcal{C}} dA_{\epsilon}(1 - v\theta_v) + \frac{1}{4\pi G}\Delta\int_{\mathcal{C}}dA_{\epsilon}[\left(v\delta\kappa_3 - \delta h\right)\delta\theta_v] \, .
  \label{eq:canonical_energy_simplified}
\end{equation}
Replacing the modified canonical energy by the entropy through Eq.~\eqref{eq:canonical_energy_dynamical_entropy_relation} gives
\begin{equation}
  \delta^2 S
  = \frac{1}{4G} \delta^2 \int_{\mathcal{C}} dA_{\epsilon}(1 - v\theta_v) + \frac{1}{2G\kappa}\int_{\mathcal{C}}dA_{\epsilon}[\left(v\delta\kappa_3 - \delta h\right)\delta\theta_v] \, .
  \label{eq:entropy_from_canonical_energy}
\end{equation}
We can now discuss Eq.~\eqref{eq:entropy_from_canonical_energy} in two cases.

\subsubsection{Case I}
\label{sec:caseI}

In this case, we assume that the null energy condition holds for the smooth two-sided one-parameter family in a neighborhood of $s=0$. Under the null energy condition, the apparent horizon lies on or inside the event horizon in the setting considered here~\cite{Hawking:1973qla,Galloway:2006ws,Kontou:2023ntd}. Since the event horizon is located at $u=0$ and the black-hole interior corresponds to $u\geqslant0$ in our Gaussian null coordinates, Eq.~\eqref{apparent horizon location} therefore implies, for each fixed $(v,x)$,
\begin{equation}
U(s;v,x)\geqslant0,\qquad
U(0;v,x)=0 \, .
\end{equation}
Here $U(0;v,x)=0$ follows because the apparent horizon coincides with the event horizon in the stationary background. Hence $U(s;v,x)$ has a local minimum at the interior point $s=0$. Fermat's theorem (interior extremum theorem) then gives
\begin{equation}
\left.\frac{\partial U(s;v,x)}{\partial s}\right|_{s=0}
=\delta U(v,x)=0 \, .
\label{eq:condition_of_delta_U}
\end{equation}
Substituting Eq.~\eqref{eq:condition_of_delta_U} into the first equation of Eq.~\eqref{eq:theta_relations}, we immediately obtain
\begin{equation}
\delta\theta_v
=\mu\delta U-D(D\delta U+\beta\delta U)
=0 \, .
\label{eq:delta_theta_v_zero}
\end{equation}
The first equation of Eq.~\eqref{eq:first_second_order_raychaudhuri}, together with Eq.~\eqref{eq:delta_theta_v_zero}, then gives
\begin{equation}
\delta T'_{ab}k^a k^b=0 \, .
\end{equation}
Since, as discussed above, $T'_{ab}$ is equivalent to $T_{ab}$ in the present analysis, this gives
\begin{equation}
\delta T_{ab}k^a k^b=0 \, .
\label{eq:deltaT_integral_zero}
\end{equation}

Substituting Eqs.~\eqref{eq:first_order_shear_contraction}, \eqref{eq:deltaT_integral_zero}, and~\eqref{eq:delta_theta_v_zero} into Eq.~\eqref{eq:canonical_energy_entropy_relation}, we obtain the following relation:
\begin{equation}
  \begin{aligned}
  {\mathcal{E}'}[\Sigma(t_2)] - {\mathcal{E}'}[\Sigma(t_1)]
    ={}& -\frac{1}{4\pi G}\int_{\mathcal H^+_{12}}
    \boldsymbol {\epsilon}^{(n-1)}
    (\xi^c\nabla_c v)\delta\sigma_{ab}\delta\sigma^{ab} \\
    &- \int_{\mathcal{H}^+_{12}}
    \boldsymbol {\epsilon}^{(n-1)}
    (\xi^c\nabla_c v)k^a k^b\delta^2T_{ab} \\
    &- \frac{1}{16\pi G}\int_{\mathcal{I}^+_{12}}
    \hat{\boldsymbol\epsilon}^{(n-1)}
    (\hat t^c\hat\nabla_c\hat u)
    \delta \hat N^{ab}\delta \hat N_{ab} \, \\
    &\leqslant 0 \, .
    \label{eq:reduced_canonical_energy_entropy_relation}
  \end{aligned}
\end{equation}
According to Eq.~(106) of~\cite{Hollands:2012sf}, the first and third terms together give a nonpositive contribution. Moreover, under the null energy condition, together with Eq.~\eqref{eq:deltaT_integral_zero} and the stationary background condition, one obtains $\delta^2 T_{ab}k^a k^b\geqslant 0$, and therefore the second term is also nonpositive. It follows that the difference of the modified canonical energy between $\Sigma(t_2)$ and $\Sigma(t_1)$ is nonpositive.

Substituting Eqs.~\eqref{eq:delta_theta_v_zero} and~\eqref{eq:condition_of_delta_U} into Eqs.~\eqref{eq:apparent_horizon_area_final} and~\eqref{eq:entropy_from_canonical_energy}, and after simplification, we finally obtain
\begin{equation}
  \delta^2 S
  = \frac{1}{4G} \delta^2 \int_{\mathcal{C}}
  dA_{\epsilon} \left(1 - v \theta_v\right)
  = \frac{1}{4G} \delta^2 A_\mathcal{T}(v) \, .
  \label{secondordervariationS}
\end{equation}
Therefore, we have shown that, under second-order perturbations of a general stationary background, dynamical black hole entropy in Einstein's gravity is proportional to the area of the apparent horizon. When $\delta U=0$ and $\delta^2 U > 0$ (so that $U>0$), the second relation in Eq.~\eqref{eq:theta_relations} implies
\begin{equation}
  \delta^2 \theta_{v}
  = \mu\delta^2 U
  - D_i(D^i\delta^2U + \beta^i\delta^2U)
  \not\equiv0\, ,
\end{equation}
since integrating over the compact horizon cross section yields
\begin{equation}
  \int_{\mathcal C}\delta^2\theta_vdA
  = \int_{\mathcal C}\mu\delta^2UdA>0\, .
\end{equation}
So the second order variation in Eq.~\eqref{secondordervariationS} is nontrivial.

In addition, from Eq.~\eqref{eq:reduced_canonical_energy_entropy_relation}, the difference of the modified canonical energy between $\Sigma(t_2)$ and $\Sigma(t_1)$ is negative; hence the dynamical black hole entropy is non-decreasing to second order in the perturbation. That is,
\begin{equation}
  \Delta\delta^2 S \geqslant 0 \, .
\end{equation}
Thus, under second-order perturbations, the dynamical black hole entropy satisfies the classical second law of black hole thermodynamics.

\subsubsection{Case II}
\label{sec:caseII}

In this case, we consider the same smooth two-sided one-parameter family in a neighborhood of $s=0$ as in \hyperref[sec:caseI]{Case I}, but no longer impose the null energy condition. Consequently, the two conditions in Eqs.~\eqref{eq:deltaT_integral_zero} and~\eqref{eq:delta_theta_v_zero} may no longer hold. Comparing Eq.~\eqref{eq:entropy_from_canonical_energy} with Eq.~\eqref{eq:apparent_horizon_area_final}, it appears that under second-order perturbations the area of the apparent horizon is not always proportional to the dynamical black hole entropy, since there remain extra terms that cannot be generally eliminated. According to~\cite{Visser:2024pwz}, there is an ambiguity in $\delta\xi^a$ or equivalently between $\delta\kappa_2$ and $\delta\kappa_3$. To ensure the area law (with respect to the apparent horizon) of the dynamical black hole entropy in the sense of second-order perturbation, $\delta\kappa_2$, $\delta\kappa_3$, and $\delta\theta_v$ have to satisfy the constraint equation
\begin{equation}
  \left(v\delta\kappa_3-\int_0^v\delta\kappa_2\,dv'\right)\delta\theta_v
  =\eta(v,x)\, ,
  \label{eq:kappa_constraint_apparent_horizon_entropy}
\end{equation}
where $\eta$ denotes
\begin{equation}
  \begin{aligned}
    \eta
    &=\frac{\kappa}{2}\Biggl\{
    \left(1+\frac{\lambda v}{\theta_u}\right)(\theta_u\delta U)^2
    -v(D\delta U)^2\theta_u
    +2\delta U(\delta\theta_u-v\partial_u\delta\theta_v)\\
    &\quad
    +(\delta U)^2(\partial_u\theta_u-v\partial_u^2\theta_v)
    -2v\delta U(D^i\delta U+\beta^i\delta U)D_i\theta_u
    \Biggr\}\, .
  \end{aligned}
\end{equation}
According to Eq.~\eqref{eq:delta_kappa2_delta_kappa3_definitions}, we can express this equation in terms of $\delta\xi^a$ as
\begin{equation}
  (vl_{[a}k_{b]}\nabla^a\delta\xi^b+\delta\xi\cdot l)\delta\theta_v
  =\eta\, .
  \label{eq:delta_xi_projection_constraint}
\end{equation}
After simplifying this relation in the chosen coordinate system, it takes the form
\begin{equation}
  \frac{v}{2}
  \left[
  \partial_u(k\cdot\delta\xi)
  -v^2\partial_v\left(\frac{\delta\xi\cdot l}{v^2}\right)
  \right]\delta\theta_v
  =\eta\, .
  \label{eq:delta_xi_nonunique_constraint}
\end{equation}

Using Eqs.~\eqref{gauge_condition_kl} and~\eqref{deltaxiintegral}, the general smooth local solution of Eq.~\eqref{eq:delta_xi_nonunique_constraint} takes the form
\begin{equation}
\delta\xi^a
=vW(u,v,x)k^a
+u\left[v\partial_vW(u,v,x)-W(u,v,x)-2X(v,x)+uY(u,v,x)\right]l^a
+uvZ^i(u,v,x)m_i^a\, .
\label{eq:caseII_general_delta_xi}
\end{equation}
Here $W$ and $Y$ are arbitrary smooth scalar functions, $Z^i$ are arbitrary smooth transverse components, and $X(v,x)$ is a smooth function on the horizon satisfying
\begin{equation}
  vX(v,x)\delta\theta_v(v,x)=\eta(v,x)\, .
\label{eq:caseII_X_constraint}
\end{equation}
The factors $v$, $u$, and $uv$ ensure that $\delta\xi^a$ is proportional to $k^a$ on $\mathcal H^+$, proportional to $l^a$ on $\mathcal H^-$, and vanishes at the bifurcation surface. Using Eq.~\eqref{eq:delta_kappa2_delta_kappa3_definitions}, we obtain
\begin{equation}
  \delta\kappa_2=W+v\partial_vW,
  \qquad
  \delta\kappa_3=W+X,
  \qquad
  \int_0^v\delta\kappa_2\,dv'=vW,
\end{equation}
With $W$ and $\partial_vW$ evaluated at $u=0$, one has $v\delta\kappa_3-\int_0^v\delta\kappa_2,dv'=vX$, and Eq.~\eqref{eq:caseII_X_constraint} follows. Thus, wherever $v\delta\theta_v\neq0$, $X=\eta/(v\delta\theta_v)$; when $v\delta\theta_v=0$, a smooth solution requires $\eta$ to vanish with the same order so that $X$ remains finite and smooth. The function $W$ remains arbitrary, while $Y$ and $Z^i$ represent off-horizon freedom, so $\delta\xi^a$ is not uniquely determined.

Even if one can choose certain vector fields $\delta\xi^a$ satisfying Eq.~\eqref{eq:delta_xi_projection_constraint} such that the dynamical black hole entropy remains proportional to the apparent-horizon area under second-order perturbations, this does not by itself guarantee that the entropy is nondecreasing at second order. Consequently, it also does not guarantee that the dynamical black hole entropy satisfies the classical second law of black hole thermodynamics.

As an additional observation, one may also consider a one-sided perturbative family, with either $s\geqslant0$ or $s\leqslant0$, while retaining the null energy condition and the other physical assumptions of \hyperref[sec:caseI]{Case I}. Since $U(s;v,x)\geqslant0$ and $U(0;v,x)=0$, one has $\delta U\geqslant0$ for $s\geqslant0$ and $\delta U\leqslant0$ for $s\leqslant0$. For a typical field theory, the stress-energy tensor is quadratic in the matter-field perturbation ~\cite{Visser:2024pwz}. Therefore, if the matter-field perturbation is counted in the usual way as $\mathcal O(s)$, its contribution to the stress-energy tensor starts at $\mathcal O(s^2)$, and hence the first-order matter flux vanishes, consistent with Eq.~\eqref{eq:deltaT_integral_zero}. A nonzero first-order matter flux would instead require the special counting in which the matter-field perturbation is taken to be $\mathcal O(|s|^{1/2})$~\cite{Visser:2024pwz}. As in \hyperref[sec:caseI]{Case I}, the first equation of Eq.~\eqref{eq:first_second_order_raychaudhuri} gives $\partial_v\delta\theta_v=0$. Since the black hole approaches a stationary state at late times, we impose $\delta\theta_v\to0$ as $v\to\infty$; hence $\delta\theta_v=0$. Integrating the first equation of Eq.~\eqref{eq:theta_relations} over the compact horizon cross section then yields
\begin{equation}
  \int_{\mathcal C}dA_{\epsilon}\,\delta\theta_v
  =\int_{\mathcal C}dA_{\epsilon}\,\mu\,\delta U
  =0\, ,
\end{equation}
where the total-divergence term vanishes. Since $\mu>0$ and $\delta U$ has a definite sign on either one-sided branch, it follows that $\delta U=0$. The remaining argument is therefore identical to that in \hyperref[sec:caseI]{Case I}, giving the same conclusions. We emphasize that this is only a possible discussion of the one-sided perturbation case. The two situations should not be regarded as equivalent, since we have not provided a complete proof of their equivalence.

\section{Conclusion and discussion}
\label{sec:conclusion}

In this work, we investigated the entropy of dynamical black holes under second-order perturbations of a stationary black-hole background admitting a bifurcate Killing horizon. Working in Gaussian null coordinates, we constructed the apparent horizon perturbatively as a deformation of the event horizon and analyzed its geometric properties up to second order, including the associated null expansions and the area element. This framework allowed us to express the second-order variation of the apparent-horizon area entirely in terms of horizon quantities defined on the background event horizon.

To establish the relation between horizon dynamics and thermodynamic quantities, we employed the covariant phase space formalism and introduced a modified canonical energy that incorporates the contribution of external matter fields. This construction leads to a balance law relating the second-order variation of the dynamical black-hole entropy to the flux entering the horizon. By combining the geometric relations derived for the apparent horizon with the Raychaudhuri equation and the horizon boundary conditions, we obtained a significant simplification of the entropy formula. In particular, under the null energy condition, all additional contributions vanish so that the entropy can be expressed solely in terms of the apparent-horizon area.

The main result of this work is that, when the null energy condition is satisfied, under second-order perturbations of a general stationary background in Einstein gravity, the dynamical black hole entropy is given precisely by the area of the apparent horizon cross section. This provides a clear and explicit realization of the entropy proposal in a near-equilibrium setting, and provides further evidence for the idea that the apparent horizon, rather than the event horizon, captures the physically relevant notion of entropy in dynamical situations. In this sense, our result strengthens the connection between horizon geometry and thermodynamic behavior beyond strict stationarity.

Several open questions remain. First, the derivation of the apparent-horizon area law relies on a consequence of the null energy condition. It would therefore be interesting to determine whether the area law survives, possibly in an altered form, when the null energy condition is violated. Clarifying these issues would be important for understanding the universality of dynamical black-hole entropy in the presence of quantum matter fields, where violations of the classical null energy condition can naturally arise and quantum corrections may break the relation between entropy and horizon area~\cite{Wall:2011hj,Bousso:2015mna}. Second, the uniqueness of the dynamical entropy beyond linear order remains to be clarified~\cite{Hollands:2024vbe}. The entropy proposal is based on the existence of a local and covariant form $\boldsymbol{B}_{\mathcal H^+}$ satisfying $\underline{\boldsymbol\Theta}(g,\delta g)=\delta\boldsymbol B_{\mathcal{H}^+}(g)$ for first-order perturbations of a stationary black hole. However, this condition does not uniquely determine $\boldsymbol{B}_{\mathcal H^+}$ beyond linear order when considering more general theories of gravity. Additional terms whose first variation vanishes on a stationary background can be added without affecting the entropy at first order, while potentially contributing to the entropy at second order. It would therefore be interesting to understand whether further physical requirements can uniquely determine the corresponding entropy functional. In particular, one may ask whether the validity of a second law beyond linear order can serve as a criterion for selecting a preferred choice of $\boldsymbol{B}_{\mathcal H^+}$ and hence a preferred notion of dynamical black-hole entropy.

Beyond these issues, it is also important to understand to what extent the present result can be generalized. Our analysis was restricted to asymptotically flat black holes in Einstein gravity. It remains unclear whether the same apparent-horizon characterization of entropy remains valid in more general settings. For example, asymptotically AdS spacetimes and higher-curvature theories provide natural testing grounds. Determining whether the dynamical entropy can still be associated with the apparent horizon and identifying the corresponding entropy functional when the Wald entropy differs from the area would offer valuable insight into the universality of dynamical black-hole thermodynamics.

\section*{Acknowledgement}
This work is supported in part by the National Natural Science Foundation of China with grants No. 12475063, No. 12075232 and No. 12247103.

\bibliography{mainRef}
\bibliographystyle{apsrev4-1}

\end{document}